\documentclass[aps,amsmath,amssymb,twocolumn,superscriptaddress,nofootinbib]{revtex4-1}

\usepackage{graphicx}
\usepackage{dcolumn}
\usepackage{bm}
\usepackage{color}
\usepackage[colorlinks=true,pdfstartview=FitV,linkcolor=blue,citecolor=blue,urlcolor=blue]{hyperref}
\usepackage[mathlines]{lineno}
\usepackage[dvipsnames]{xcolor}
\usepackage[utf8]{inputenc}
\usepackage[T1]{fontenc}
\usepackage{newtxtext,newtxmath}
\usepackage[dvipsnames]{xcolor}
\usepackage{multirow}
\usepackage[caption=false]{subfig}
\usepackage{tabularx}
\usepackage{siunitx}
\usepackage{float}
\usepackage{braket}
\usepackage{amsmath}
\usepackage{gensymb}

\usepackage{soul}

\newcommand{\etaph}{\mbox{$\eta_{\rm ph}$}}
\newcommand{\Edep}{\mbox{$E_{\rm dep}$}}
\newcommand{\Eabs}{\mbox{$E_{\rm abs}$}}
\newcommand{\nqp}{\mbox{$n_{\rm qp}$}}
\newcommand{\nought}{\mbox{$n_{0}$}}
\newcommand{\tqp}{\mbox{$\tau_{\rm qp}$}}

\newcommand{\dnqp}{\mbox{$\delta n_{\rm qp}$}}

\newcommand{\Nqpr}{\mbox{$N_{\rm qp}^r$}}
\newcommand{\gin}{\mbox{$\Gamma_{\rm in}$}}
\newcommand{\gout}{\mbox{$\Gamma_{\rm out}$}}

\newcommand{\bigo}[1]{$\mathcal{O}({#1})$}
\newcommand{\um}{\mbox{$\mu {\rm m}$}}
\newcommand{\umthree}{\mbox{$\mu {\rm m}^{3}$}}
\newcommand{\umminusthree}{\mbox{$\mu {\rm m}^{-3}$}}

\begin{document}

\title{Quantum Parity Detectors: a qubit based particle detection scheme with meV thresholds for rare-event searches}

\author{K. Ramanathan}
\email{karthikr@wustl.edu}
\affiliation{Department of Physics, Washington University in St. Louis, St. Louis, MO, 63130, USA}

\author{B.~J.~Sandoval} \affiliation{Department of Physics, California Institute of Technology, Pasadena, CA, 91125, USA}
 
\author{J.~E.~Parker} \affiliation{Department of Physics, California Institute of Technology, Pasadena, CA, 91125, USA}

\author{L.~M.~Joshi} \affiliation{Department of Condensed Matter Physics, Weizmann Institute of Science, Rehovot, Israel}

\author{A.~D.~Beyer} \affiliation{Jet Propulsion Laboratory, California Institute of Technology, Pasadena, CA 91109, USA} 

\author{P.~M.~Echternach} \affiliation{Jet Propulsion Laboratory, California Institute of Technology, Pasadena, CA 91109, USA} 

\author{S.~Rosenblum} \affiliation{Department of Condensed Matter Physics, Weizmann Institute of Science, Rehovot, Israel}

\author{S~R.~Golwala} \affiliation{Department of Physics, California Institute of Technology, Pasadena, CA, 91125, USA}

\date{\today}

\begin{abstract}
The next generation of rare-event searches, such as those aimed at determining the nature of particle dark matter or in measuring fundamental neutrino properties, will benefit from particle detectors with thresholds at the meV scale, 100\textendash1000$\times$ lower than currently available. Quantum parity detectors (QPDs) are a class of proposed quantum devices, extending recent work on superconducting qubit sensors, that exploit the fingerprints of single quasiparticle tunneling across a coherent weak-link as their detection concept. As envisioned, phonons generated by particle interactions within a crystalline substrate cause an eventual quasiparticle cascade within a surface-patterned superconducting qubit element. This process alters the fundamental charge parity of the device in a binary manner, which can be used to deduce the initial properties of the energy deposition. Novelly, this work lays out multiple resonator coupled readout schemes depending on qubit architecture, provides an analytic formulation for reconstructing sensor energies, and details strategies for multiplexing large arrays of sensors. We further compute the sensitivity of QPDs and detail an R\&D pathway to demonstrating sub-eV energy deposit thresholds. \end{abstract}

\maketitle

\section{Introduction} \label{sec:introduction}

A key science requirement for the next generation of rare-event searches, such as those looking for very light particle dark matter or expanding neutrino measurement regimes \cite{{battaglieri2017us, *kolb2018basic, *fleming2019basic, papoulias2019recent}}, is the ability to identify energy deposits $\ll$ eV within a macroscopic ($\mathcal{O}$(gm)) target mass. For example, theoretically well-motivated particle dark matter candidates with $\mathcal{O}$(10\textendash10$^4$)~keV$c^{-2}$ masses may interact with an atomic target and kinematically transfer only $\mathcal{O}$(1-10$^3$)~meV \cite{essig2022snowmass2021}. Further complicating this picture is that, as the interaction energy (momentum) scale drops below the $\sim$eV ($\sim$keV$c^{-1}$) level in common crystalline target materials, it is predominantly only collective excitations that are created, the most common of which are phonons (lattice vibration quanta) \cite{trickle2022direct}. In the case of a dark matter candidate coupled to the Standard Model through an electromagnetic current (e.g. dark photons), interactions with an ionic crystal can likewise produce optical phonons with energies down to $\mathcal{O}$(10-10$^2$) meV \cite{trickle2020MultiChannel}.  Transition Edge Sensors (TESs) \cite{irwin1995quasiparticle} have demonstrated the lowest thresholds for detection of phonons, at the $\mathcal{O}$(100)~meV to eV level \cite{rothe2018tes,alkhatib2021light,fink2020characterizing,tesseract2023results, anthony2025low}, and Kinetic Inductance Detectors (KIDs) are on a similarly trajectory \cite{wen2022performance, moore2012position, delicato2023low}. However, it remains an outstanding particle physics community priority to demonstrate \textit{single-phonon-sensitive detectors} as tools in the search for new physics \cite{essig2022snowmass2021}.

Concurrently, the past decades have seen rapid progress in the field of superconducting qubits for Quantum Information Science (QIS) \cite{nakamura1999coherent,somoroff2023millisecond}. These computing building blocks have been shown to be affected by environmental radioactivity, producing correlated errors across an entire wafer \cite{wilen2021correlated}, induced by charge and phonons produced within the substrate. An active area of QIS R\&D is in mitigating these effects by engineering phonon or quasiparticle absorbers and sinks, and gap engineering \cite{iaia2022phonon,bargerbos2023mitigation,martinis2021saving, mcewen2024resisting}. 

Crucially, the relevant energy scales of the physical phenomena affecting qubit coherence is matched to the particle physics requirements discussed above. For example, in a common qubit material like aluminum with a superconducting Cooper pair-breaking gap of $2\Delta\approx 400\,\mu$eV, any resultant quasiparticle (broken Cooper-pair) from phonon absorption can tunnel across the Josephson Junction at the heart of the circuit. This tunneling process has a fingerprint, modifying the energy level structure of the qubit and  possibly decohering it \cite{shaw2008kinetics,catelani2014parity, catelani2011relaxation}.

Quantum Capacitance Detectors (QCDs) \cite{shaw2009qcd, bueno2010proof}, based on Single Cooper-Pair Box (CPB) qubit designs \cite{nakamura1999coherent}, exploit this quasiparticle tunneling process for far-IR radiation detection, and they have since demonstrated the ability to count 1.5~THz photons \cite{echternach2018single} using an RF-based readout scheme that lends itself to easy multiplexing, with over 400 QCDs chained on a single feedline \cite{echternach2021large}. It is natural to consider whether such sensors can be repurposed as generic phonon-mediated detectors 
for rare-event searches. Complementary work by Linehan et al. \cite{linehan2024estimating} has investigated the efficacy of using correlated errors in conventional transmon qubits to build phonon sensitive detectors, but concludes that energy resolutions are limited to the $\mathcal{O}$(10)~eV level. The SQUAT concept of Fink et al. \cite{fink2023superconducting} dramatically re-imagines a qubit's architecture, directly coupling it to the RF feedline, while still measuring changes to charge parity as with the seminal QCD work. The SQUAT scheme, while promising, requires coupled sensor/readout design, use of quantum-limited amplifiers, and tradeoffs in sensor bandwidth \& SNR. Further, the direct feedline coupling is potentially susceptible to an aggravated excess quasiparticle density, arising from hypothesized thermal or multi-photon processes observed in qubits \cite{chowdhury2025theory} and resonators \cite{de2012microwave, goldie2012non}.

\begin{figure}[!ht]
    \includegraphics[width=0.48\textwidth]{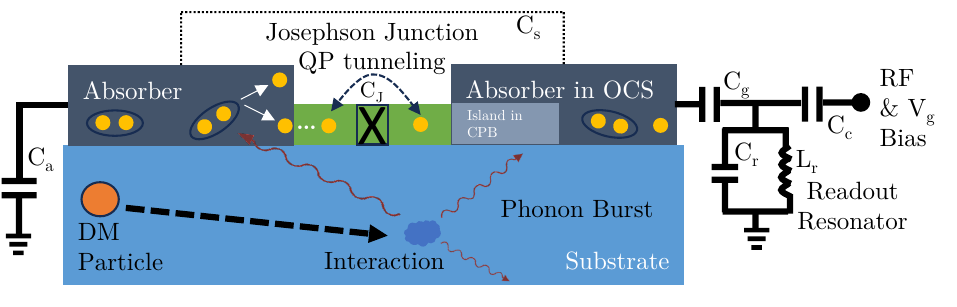}
    \caption{\label{fig:operation} Schematic of the operating principle of both QPD styles. A quasiparticle trapping region containing the Josephson junction (green) is sandwiched either by large superconducting absorbers (OCS, gray) or an absorber and small island (CPB, light gray). The structure is embedded in a resonant readout circuit, with identified circuit terms, and patterned on the surface of a crystalline substrate (blue).}
\end{figure}

\begin{figure*}[!ht]
    \includegraphics[width=0.98\textwidth]{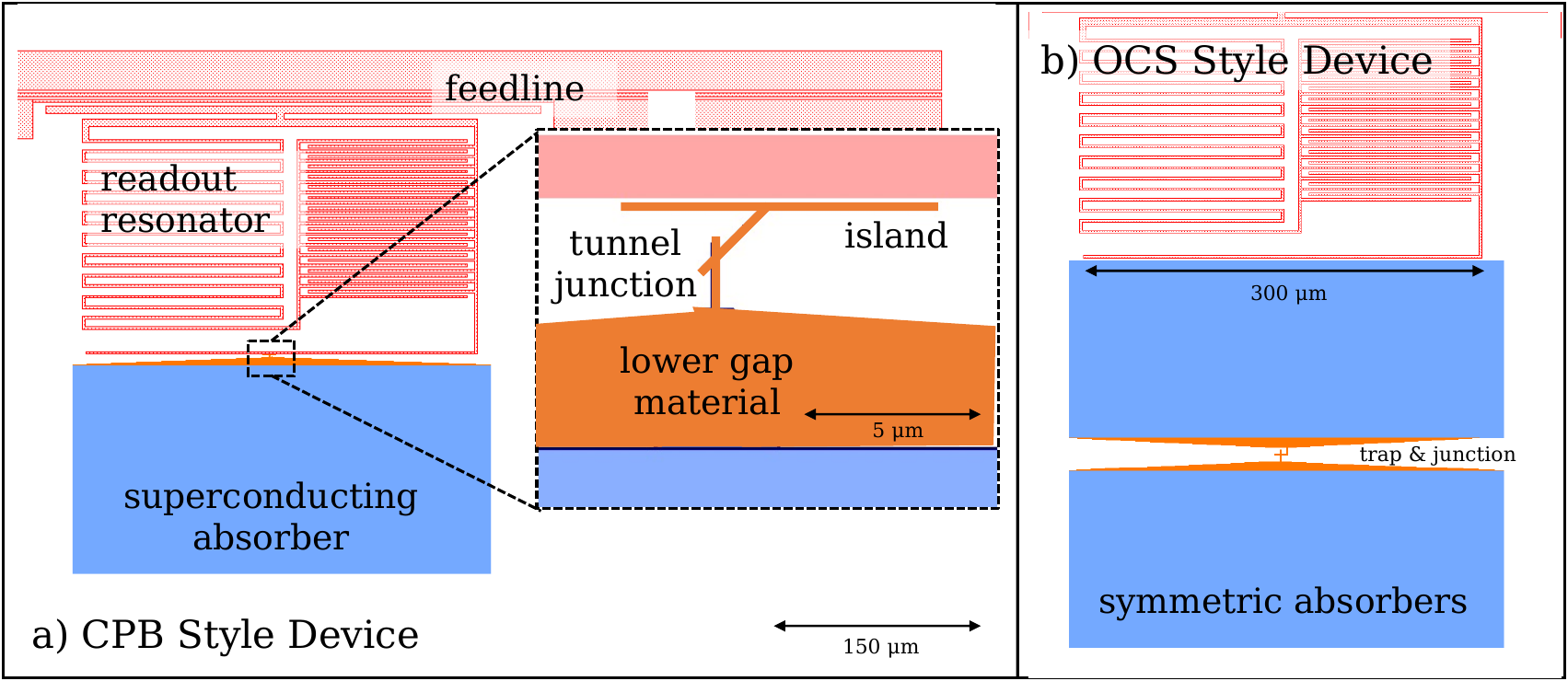}
    \caption{\label{fig:layout} Stylized mask layouts of QPDs. Items in red are made of a high gap material (e.g. Nb or Ta), used for the feedline and resonator. Blue represents an intermediate gap material (e.g. Al) for the phonon absorber. Orange highlights the low gap material (e.g. Hf) trap and junction. a) QCD derived sensor, with large absorber connected to small island. We note the potential use of a lower  gap (than the absorber) material to connect the large absorber to the tunnel junction (and the island) to enable quasiparticle trapping and multiplication. b) Resonator and absorber region for an OCS design, with mostly identical components except larger, symmetric pads that introduce a shunt capacitance. Trapping may be implemented on both sides of the junction.}
\end{figure*}

This note lays out the intermediate approach of a resonator coupled charge parity sensitive pair-breaking detector we term Quantum Parity Detectors (QPDs). This architecture considers phonon sensitive CPB designs as well as Offset Charge Sensitive (OCS) transmons \cite{serniak2018hot, koch2007charge}, platforms that are well studied in the literature. Sec. \ref{sec:concept} will outline the detector concept. Sec. \ref{sec:devicephysics} will describe relevant physical principles along with the readout mechanism, including providing for an analytical formulation of tunneling rates that allows for direct determination of device sensitivity. Sec. \ref{sec:deviceperformance} will detail the estimated performance along with cataloging expected noise sources. Finally, Sec. \ref{sec:additional} will outline future choices and challenges to be addressed.

\section{Device Concept} \label{sec:concept}

A schematic overview of the device concept is given in Fig. \ref{fig:operation}, following well established phonon mediated quasiparticle trapping designs such as \cite{irwin1995quasiparticle}. A rare-event interaction such as that of a dark matter particle scattering within the substrate creates an energy deposit $E_{\rm dep}$ that will result in the production and propagation of phonons. The substrate is a high-resistivity, macroscopically thick crystal, typically $\mathcal{O}$(gm) in mass, with silicon, germanium, and sapphire (Al$_2$O$_3$) being common choices for rare-event experiment architectures \cite{knapen2018detection,essig2016direct}. The QPD portion consists of two superconducting pads patterned on the substrate surface coupled by a Josephson Junction of a $\mathcal{O}$(100)~nm side lengths \cite{shnirman1997quantum}, akin to a traditional qubit design \cite{gao2021practical, koch2007charge}. Depending on whether it is an OCS or QCD-style sensor, with heritage from transmon or CPB qubits respectively, the pads can be symmetrically large (\textit{absorber-absorber}) or asymmetric (\textit{absorber-island}). The produced phonons impinge on and can be absorbed by the pads, breaking Cooper-pairs and creating quasiparticles. These absorber quasiparticles diffuse to and are trapped in an overlapping smaller volume, lower superconducting gap region called the trap (also known as the junction leads in qubit specific literature), and dramatically change the quasiparticle density near the Josephson Junction. This modulation of quasiparticle density, $\delta n_{\rm qp}$, and subsequent rapid tunneling back-and-forth across the junction, can be sensed via a quantum-mechanical interaction with the coupled readout resonator of bare frequency $\omega_r$ ($=1/\sqrt{L_r C_r}=2 \pi f_r$). The absorber is made of $\mathcal{O}$(100s)~nm (e.g. 500~nm), relatively thick aluminum (due to its capability to absorb phonons and diffuse quasiparticle efficiently), while the junction is to be made of a $\mathcal{O}$(100)~nm thick, low gap material (such as hafnium) for the purpose of efficient quasiparticle trapping \cite{booth1987quasiparticle,goldie1990quasiparticle}. Hafnium is a promising candidate as tunnel junctions have been successfully demonstrated in the literature \cite{kim2012development,kraft1998use}, though not as part of a qubit architecture. 

Proposed layouts of the two types of QPDs can be seen in Fig. \ref{fig:layout}, highlighting the discussed elements of the absorber, island, resonator, readout feedline, and Josephson Junction.

\begin{figure}[!h]
    \includegraphics[width=0.48\textwidth]{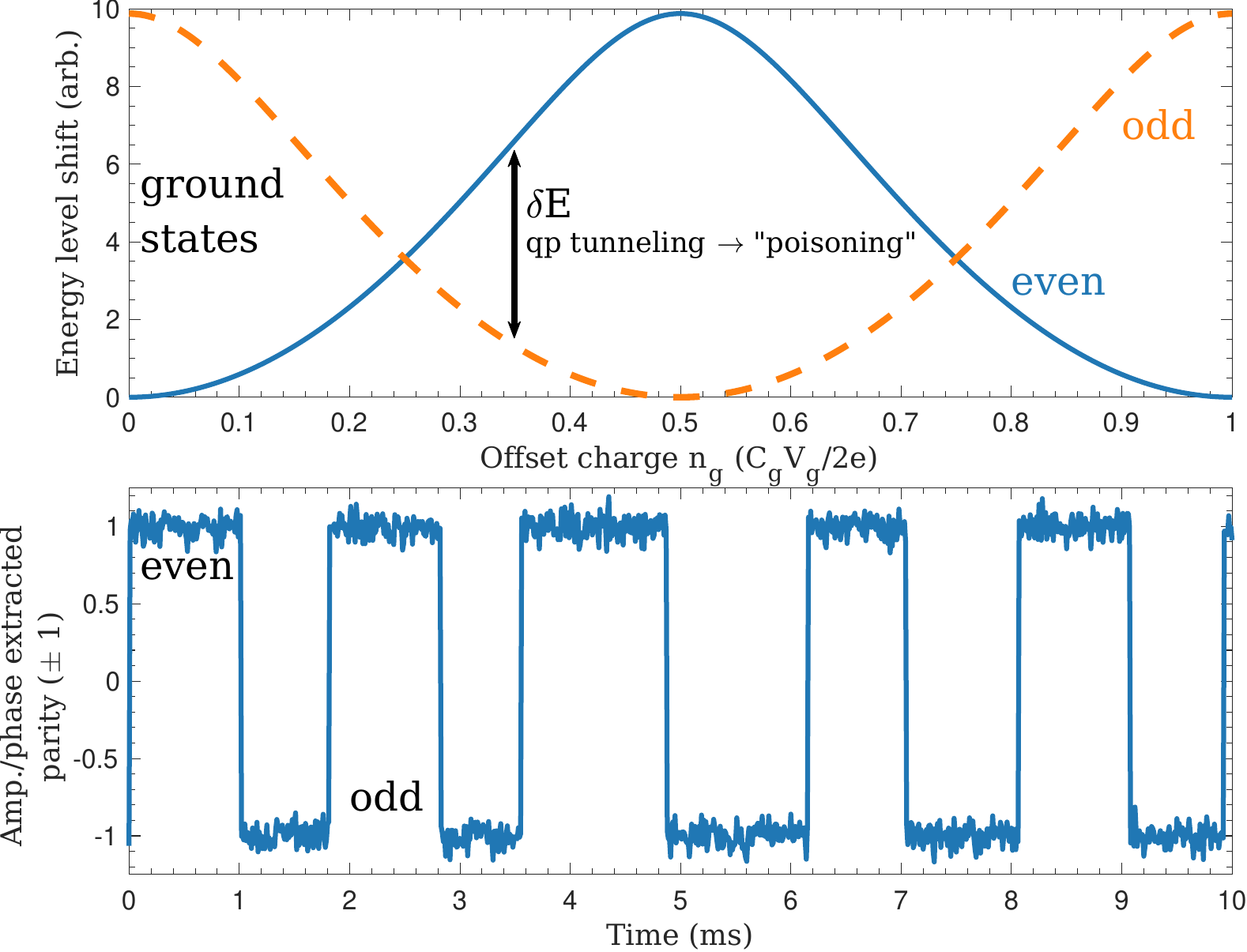}
    \caption{\label{fig:stylized} \textit{Top:} Stylized ground-state energy level shifts for a charge qubit, showing the different curves for even and odd number of charges on the island, with $\delta E$ difference in energy. A quasiparticle tunneling event shifts the qubit from one state to the other. \textit{Bottom: } Simulated resonator telegraph signal from background quasiparticle tunneling events. Each tunneling event causes a charge parity shift and thus an amplitude/phase change in the resonator signal, and this rate is proportional to the quasiparticle density within the absorbers.}
\end{figure}

Charge qubits (which includes CPBs and transmons) are superconducting circuits engineered to have addressable ``charge" basis states \textemdash\ simplistically, their properties are determined by the number of Cooper pairs present on an isolated island \cite{wallraff2004strong,nakamura1999coherent,koch2007charge}. As will be elucidated in Sec. \ref{sec:qpd}, charge qubits can be treated as anharmonic quantum oscillators, reducible to a two-level quantum system, and thus have found great use as building blocks for quantum computing \cite{you2005circuits,krantz2019quantumengineer}. However, due to their environmental sensitivity, these schemes can be affected by the incoherent tunneling of quasiparticles across the Josephson Junction. The ground state energy of the qubit is actually a sinusoidal function of the offset charge, $n_g$, on the island, as shown in Fig. \ref{fig:stylized} Top. The addition of a single quasiparticle can be interpreted as a change in $n_g$ by 0.5, causing the qubit to switch between the ``odd" and ``even" curves, with a difference in energy of $\delta E$. Hence, it is said that addition or removal of a quasiparticle changes the ``charge parity" of the system \cite{serniak2018hot}. Even in the case of very low temperature operation, with a thermally suppressed quasiparticle population, non-equilibrium quasiparticles from other sources pose this risk of quasiparticle poisoning \cite{aumentado2004nonequilibrium}, potentially ruining quantum coherence. Regardless of QPD type, a microwave resonator with bare resonant frequency $\omega_r$ can be capacitively coupled to the qubit to monitor the charge parity\footnote{In the CPB case, the parity shift is interpreted as a change in the device capacitance (hence Quantum Capacitance Detector)}. Any excess population of non-equilibrium quasiparticles created by phonon absorption can tunnel across the junction and thus be sensed by each tunneling event's impact on $\omega_r$, with the tunneling rate ($\mathcal{O}(1)$ kHz$\cdot \mu$m$^{3}$) proportional to this quasiparticle density. The excess rate of charge parity transitions should thus be proportional to the original substrate energy deposition, providing a basis for using QPDs for particle detection \cite{echternach2018single}. Fig. \ref{fig:stylized} Bottom shows an example of the expected resonator response to charge parity shifts, called a telegraph signal. We consider even charge parity to be zero or an even number of quasiparticles (with odd being the complement) to have tunneled across the barrier. Additionally, an applied voltage bias $V_g$, adjusted with the aid of a gate capacitor $C_g$, allows for tuning of the gate charge operating point to maximize the observed charge parity switch signal.  

In the QPD scheme, the underlying qubits are always operated in their ground state, requiring no complex state manipulation. Additionally, the sensors do not interact with each other. Thus, QPDs have the added benefit of easy multiplexability, as each qubit can be coupled to a readout resonator of a slightly different resonant frequency. The expected maximal dispersive shift of a GHz resonator due to a charge parity switching event will be $\mathcal{O}$(100)~kHz for OCS style devices and $\mathcal{O}$(10) MHz for CPB style devices. See Sec. \ref{sec:readout} for more details on the expected parity frequency shift. Furthermore, even the lower bound parity frequency shift is a significant fraction of the resonators FWHM $f_r/Q$ for a loaded quality factor $Q_r\sim10^4$. The frequency noise (while in either the even or odd states) is nominally expected to be $\mathcal{O}(100)$~Hz \cite{burnett2018noise,ramanathan2024significant}, hence we expect the even and odd states to be easily resolvable. In other words, any reasonable RF probe will be able to continuously distinguish even vs. odd states. There is however the added complexity of having to tune the gate charges for each QPD, due to differences between QPDs arising from charge inhomogeneities and associated drifts. For a single QPD this problem can be overcome by actively monitoring $n_g$ via the size of observed charge parity shifts (or other related variables, such as changes in the qubit transition frequency) and adjusting the voltage gate bias, a technique already used in qubit systems \cite{connolly2023coexistence}. For multiplexed systems, we may continuously sweep the gate bias voltage through one full period of gate charge. A tunneling event will then interrupt the periodic output signal differently depending on whether the device utilizes an OCS or CPB qubit. A full discussion of the expected response when sweeping the gate voltage is given in Section \ref{sec:readout}.

\section{Device Physics} \label{sec:devicephysics}
\begin{figure}[!t]
    \includegraphics[width=0.48\textwidth]{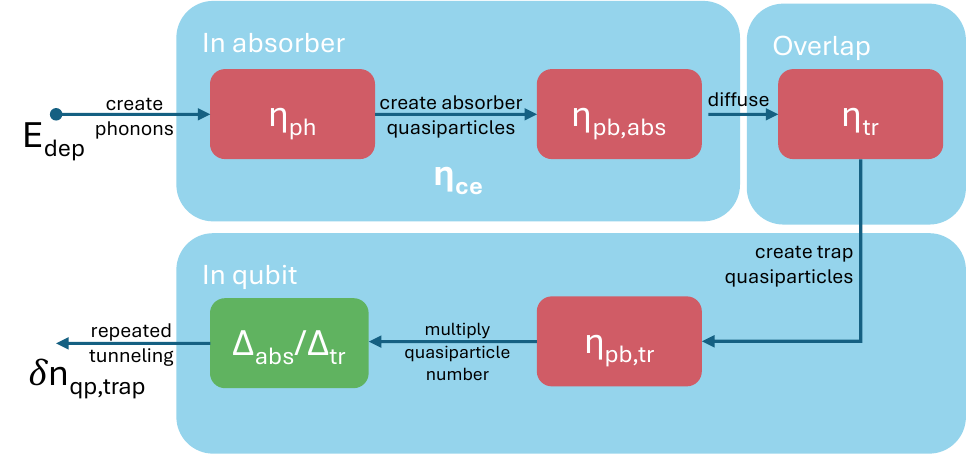}
    \caption{\label{fig:effstring} Stylized chart showing the string of efficiencies that modulate the initial substrate deposit $E_{\rm dep}$ into a changing quasiparticle density, $\delta n_{\rm qp,trap}$ near the junction region of the QPD. The first 4 efficiencies reduce either the available energy or quasiparticle population to be sensed, while the final act of trapping induces a quasiparticle multiplication. Specific details of absorber efficiencies can be found in Sec. \ref{sec:phonon}, while quasiparticle physics in the overlap and trap region is touched on in Sec. \ref{sec:quasiparticle}}
\end{figure}

An energy deposition into the substrate generates a phonon event whose absorption into a superconducting film on the surface engenders a pulse-like time-varying change in the quasiparticle density in the absorber $\delta n_{\rm qp,abs}(t)$. Nominally, there are multiple time-scales that need to be considered for both phonons and quasiparticles, including cyclic transfer of energy between both channels \cite{kaplan1976quasiparticle}. In tandem, there are a string of efficiencies modulating the transfer of energy between each step of the overall process, for example factoring in phonon losses to surface-mediated down-conversion or efficiency of quasiparticle trapping into the trapping region. Various references have cataloged said efficiencies and time-scales \cite{pyle2012optimizing, fink2022gram} but crucially there is no singular literature model that reproduces observed pulse shapes across various detector modalities. Rather it seems experiment-specific factors of material selection, geometry, mounting, film properties, application of a bias field, etc. all impact the detector response and must be disentangled empirically. As such, we limit ourselves to simplified models for determining $\delta n_{\rm qp,abs}$ and $\delta n_{\rm qp,trap}$ (the quasiparticle density in the trap), useful to get a flavor of detector operation. We first consider the expected efficiency of capturing phonon energy by the absorber pads, then follow with a discussion on the expected efficiency of quasiparticle number within the trap, visually summarized in Fig. \ref{fig:effstring}.

\subsection{Phonon Response \& Efficiencies} \label{sec:phonon}

Following the description in Ref. \cite{saab2002search}, the initial burst of phonons will be around the Debye frequency of $\mathcal{O}$(10)~THz. Decay and scattering processes cause these relatively high-frequency phonons to downconvert within a few $\mu$s to phonons that propagate quasi-diffusively with a speed less than the crystal's sound speed. Once the phonon energies are low enough (still meV and thus athermal for a sub-Kelvin substrate), the mean free path exceeds the crystal dimensions. They travel ballistically at the sound speed and are subject primarily to reflections at the bare crystal surfaces or absorption at interfaces between the crystal and other materials (thin films and mounting points). Typical optical branch phonon energies in commonly considered crystalline substrates are of $\mathcal{O}$(10-100)~meV with the high tail of acoustic phonons ($k_BT_{\rm debye})$ reaching this level. 

To begin understanding the transfers of energy within the detector, we first model the fraction of the deposited phonon energy $E_{\rm dep}$ that enters the absorber, called the phonon collection efficiency $\eta_{\rm ph}$ in some literature (e.g Ref.~\cite{ren2021design}). Using the prescription Ref.~\cite{golwala2022novel} sketches out, $\eta_{\rm ph}$ can be maximized via the heuristic,
\begin{equation} \label{eq:abs}
    \eta_{\rm ph} \sim f_{\rm surf} f_{\rm abs} N_{\rm surf}(1-f_{\rm loss}) \leq 1.
\end{equation}
$f_{\rm surf}$ is the instrumented fraction of the substrate's surface, $f_{\rm abs}$ is the probability of absorption per interaction with the surface sensors, $N_{\rm surf}$ is the number of phonon surface encounters before thermalization below the absorber gap, and $f_{\rm loss}$ is a parasitic loss term (e.g. from phonon downconversion and losses to chip mounting). Average phonon sound speeds of km$\cdot$s$^{-1}$ and nominal lifetimes against anharmonic downconversion of $\mathcal{O}$(1)~ms, suggest $N_{\rm surf}>10^3$ for mm thick substrates. With $f_{\rm abs} > 0.1$ for $\mathcal{O}$(100)~nm thick aluminum on silicon (Al has a 1.6~$\mu$m pair-breaking length \cite{kaplan1976quasiparticle} and sound-speed mismatch limits the phonon transmittance to $<0.9$), $f_{\rm surf}$ as low as a few \% may, optimistically, be sufficient for complete energy collection. Work by \cite{ren2021design} provides evidence that $\eta_{\rm ph}>0.95$ is achievable in a phonon-mediated detector context though with almost complete surface coverage, while recent results from Ref. \cite{chang2025first} detail $\eta_{\rm ph}\sim0.3$ overall with a low 1\% surface coverage, both for Si substrates. These results are broadly in line with the presented formalism, though care must be taken to properly understand the $f_{\rm loss}$ term across different architectures.

As Fig. \ref{fig:layout} shows in light red, our QPD designs have feedline and resonator structures that may prove lossy to phonons. Parasitic losses to these inactive/passive surface metal films (``dead metal") and mountings must not unintentionally degrade the phonon population and lifetime. The exact form of $f_{\rm loss}$ is an open question, but recent work by \cite{wen2025strategic} suggests that dead metal losses may scale with the \textit{volume} of the loss channel and that a material like niobium, commonly used to pattern secondary planar structures, is about $\sim$10~\% as efficient as aluminum in absorbing Si substrate athermal phonons. As such, one can plausibly model $f_{\rm loss} \sim \Sigma_i \xi_{\rm i} V_{\rm i}/V_{\rm metal}$ as a sum over all loss channels with loss volumes $V_{\rm i}$ (an effective volume when modeling losses to the mounting), total surface metal volume $V_{\rm metal}$, and a relative lossiness measure $\xi_{i}$. Additionally, the substrate-absorber transmittance must be high enough to not degrade $f_{\rm abs}$. 

Once the phonons have entered the film, the next source of energy loss is incurred as a finite fraction of the phonon energy in the absorber is converted into quasiparticles (due to quasiparticles emitting sub-$\Delta_{\rm abs}$ (the superconducting gap of the absorber) phonons as they relax to the gap after creation). This phonon-to-quasiparticle pair-breaking efficiency in the absorber is parametrized by $\eta_{\rm pb,abs}$. Simulation work of \textit{photon} absorption in a thin-film superconductor suggests a general range of $\eta_{\rm pb}\sim0.4-0.6$ \cite{guruswamy2014quasiparticle,kozorezov2000quasiparticle}. It is reasonable to assume a similar $\eta_{\rm pb,abs}$ for phonon absorption.

We can combine the phonon absorption and pair-breaking into a single dimensionless \textit{total energy collection efficiency}\footnote{in some literature, e.g. Ref.~\cite{temples2024performance}, this is itself called the phonon collection efficiency} \cite{moore2012position},
\begin{equation}\label{eq:etaph}
    \eta_{\rm ce} = \eta_{\rm pb,abs}\eta_{\rm ph} = N_{\rm sens}V_{\rm abs} \Delta_{\rm abs} \delta n_{\rm qp,abs} / E_{\rm dep}
\end{equation}
that helps directly translate from deposited energy $E_{\rm dep}$ to the produced quasiparticle number $V_{\rm abs} \delta n_{\rm qp,abs}$ (for absorber volume $V_{\rm abs}$ and quasiparticle density $\delta n_{\rm qp,abs}$). $N_{\rm sens}$ is the number of sensors populating the surface and maps to changes in $f_{\rm surf}$ from Eq. \ref{eq:abs}, assuming equal phonon energy splitting.

To quantify phonon absorption for a representative QPD design, we examine the case of a 1~cm $\times$ 1~cm $\times$ 0.1~cm Si chip. A cm-long minimal niobium Coplanar Waveguide (CPW) feedline\footnote{with feature sizes still able to be optically lithographed.} has an effective volume of about 35,000~\umthree. A fairly standard lumped element resonator, akin to the one shown in Fig. \ref{fig:layout}, represents a small 450 \umthree\ increase to that value per QPD (the feedline ground plane is reduced at that location to increase the coupling quality factor). Both are dwarfed, in arrays of multiple QPDs, by the expected 60,000~\umthree\ volume of a \textit{single} QPD's absorber pads. 

A design with a 4\% area coverage requires a 2.5 cm feedline and about 50 QPDs. We can estimate $\eta_{\rm ce}$ by comparing this design to Ref. \cite{temples2024performance}, which details the performance a mixed Al/Nb phonon-mediated KID. They present a raw $\eta_{\rm ce}\approx0.009$ in a mixed Al/Nb device, but their $f_{\rm surf}$ is a very small $\sim$0.0017\ and $f_{\rm loss}$ is estimated to be $>0.95$, with phonons lost to both a large volume of Nb and the mounting. Scaling $f_{\rm surf}$ to 0.04 and eliminating all the Nb except for the feedline yields a predicted $\eta_{\rm ce}$ of about 0.3, dominated by losses to the mounting ($f_{\rm loss}\sim0.35$). A similar survey of the broader literature on phonon-mediated detectors coupling sapphire, silicon, and germanium substrates to both thick ($\mathcal{O}$(100)~nm) and thin-film ($\mathcal{O}$(10)~nm) aluminum absorbers yields an estimate of $\eta_{\rm ce} \sim 0.1$\textendash0.4~\cite{moore2012position, ren2021design,pyle2009surface,cardani2021final,cruciani2022bullkid,chang2025first} for our thick-film, percent level surface coverage absorber scenario.  

Optimizing the efficiencies presented in this section will largely rely on proper engineering. For instance, novel resonator architectures with parallel-plate capacitors \cite{zotova2024control}, which have already been demonstrated for transmon readout, can possibly reduce resonator volumes down to a negligible $\sim$75~\umthree\ per QPD over the feedline volume with an accompanying small footprint. In principle, the issue of dead-metal can be rendered entirely moot by adopting a flip-chip architecture \cite{foxen2017qubit,satzinger2019simple}, where all qubit readout and control architecture is moved to a \textit{separate} chip located a few $\mu$m above the qubits and is only capacitively or inductively coupled. In addition, work by Ref. \cite{romani2024transition} has demonstrated the ability to hang chips using wirebonds, which presents an attractive strategy to mitigate any losses to the mounting. 


\subsection{Quasiparticle Response \& Efficiencies}
\label{sec:QPResponse}
Once the quasiparticles are produced within the absorber and diffuse towards the junction, they can recombine before they  become trapped in the lower gap region. Hence, we expect only a fraction of the initially generated quasiparticles to actually become trapped, parametrized by the trapping efficiency $\eta_{\rm tr}$. To estimate $\eta_{\rm tr}$, we use a 1D analytic diffusion model \cite{angloher2016quasiparticle}, described in detail in Appendix \ref{app:diffsketch}. We find that this term is expected to be $\mathcal{O}$(0.1) for $\mathcal{O}$(100 $\mu$m) absorber lengths, $\mathcal{O}$(10 $\mu$m) trap lengths, and $\mathcal{O}$(100 nm) thick films.

A quasiparticle becomes trapped once it enters the lower gap region and decays in energy to $\Delta_{\rm tr}$ (the superconducting gap of the trap) by phonon emission. Because we have $\Delta_{\rm abs} \gg \Delta_{\rm tr}$, the phonon cascade generated as the quasiparticle relaxes to the lower gap can break Cooper pairs and generate more quasiparticles, yielding a quasiparticle multiplication effect. The upper bound of this multiplication is roughly equal to $\Delta_{\rm abs}/\Delta_{\rm tr}$ \cite{moffatt2016two,booth1987quasiparticle}, where we have made the approximation that all absorber quasiparticles entering the trap have energy $\Delta_{\rm abs}$, which is reasonable given how quickly excited quasiparticles decay to near $\Delta_{\rm abs}$ by phonon emission \cite{kaplan1976quasiparticle}. In the pair-breaking processes occurring within the trap, we will once again lose a fraction of energy to sub-$\Delta_{\rm tr}$ phonons emitted as the newly created quasiparticles relax to the gap. Hence, we once again must include a pair-breaking efficiency $\eta_{\rm pb,tr}$, expected to be similar in magnitude to $\eta_{\rm pb,abs}$.

Assuming again an equal split of energy across $N_{\rm sens}$ number of sensors, we can now estimate the maximum number of quasiparticles $N^r_{\rm qp}$ produced and trapped in the lower gap region near the junction of a single QPD as
\begin{align}\label{eq:qp_max}
    N_{\rm qp}^r &= \frac{E_{\rm dep}}{N_{\rm sens}\Delta_{\rm abs}} \eta_{\rm ph} \eta_{\rm pb,abs} \eta_{\rm tr} \eta_{\rm pb, tr} \frac{\Delta_{\rm abs}}{\Delta_{\rm tr}}\notag\\ &= \frac{E_{\rm dep} \eta_{\rm ce} \eta_{\rm tr} \eta_{\rm pb,tr}}{N_{\rm sens}\Delta_{\rm tr}} = \frac{E_{\rm abs} \eta_{\rm tr} \eta_{\rm pb,tr}}{\Delta_{\rm tr}},
\end{align}
having strung together factors from our preceding discussion in Sec. \ref{sec:phonon}. We note that the decrease in quasiparticle number with increasing $N_{\rm sens}$ is in part balanced by the increase in $\eta_{\rm ce}$ with more sensors. For future convenience in Sec. \ref{sec:resolutionandthreshold}, where we discuss intrinsic sensor resolutions, we have introduced $E_{\rm abs}$ in the final expression, as the energy effectively absorbed and converted into quasiparticles within the absorbers of a single QPD.

Now that we have modeled the expected quasiparticle production within the trapping region, we proceed to construct a simplified $\delta n_{\rm qp,trap}$(t) model of the dynamic change in quasiparticle density within the trap. We introduce two time constants: a quasiparticle injection timescale $\tau_{\rm inj}$, which characterizes how long quasiparticles take to enter into the trap; and $\tau_{\rm qp}$, the recombination time constant within the film~\cite{kaplan1976quasiparticle,leman2012invited,pyle2012optimizing}. $\tau_{\rm inj}$ subsumes the expected quasiparticle lifetime within the absorber, phonon time constants related to how long $>2 \Delta_{\rm abs}$ phonons are present in the substrate to break pairs in the absorber films, and diffusion related time constants. We intuit that $\tau_{\rm inj}$ will be set by the longest of these constants. To provide a scale, literature measurements of $\tau_{\rm qp}$ range from 0.1\textendash10~ms in Al \cite{barends2008quasiparticle,de2014fluctuations,moore2012position,zmuidzinas2012superconducting,baselmans2009long}, depending on the quiescent quasiparticle density $n_0$. Qubit and related superconducting systems report $n_0\sim0.1$\textendash$1\,\mu$m$^{-3}$ (e.g. Refs.~\cite{connolly2023coexistence,saira2012vanishing,riste2013millisecond}), which would correspond to $\tau_{\rm qp}\gg$~ms for a standard generation-recombination model with Al recombination constant $R \approx 10$~\umthree$\rm{s}^{-1}$ \cite{chang2023supercdms}. We assume a vanishing rise time for the phonon population, neglecting the time it takes for the phonons produced by the energy deposition to become an approximately spatially homogeneous population in the substrate, because that rise time is only a few times the $\mathcal{O}$($\mu$s) sound travel time for the considered substrates. And as cataloged by Ref. \cite{golwala2022novel}, we expect phonon absorption lifetimes, $\tau_{\rm abs}$, of $\mathcal{O}$(ms). Overall, we can write and solve the forced differential equation of a single sensor (see Appendix \ref{app:pulse} for specifics) resulting in a pulse model of the change in quasiparticle density within the trap as:
\begin{align}\label{eq:qp_pulse}
    \delta n_{\rm qp,trap}(t) &= \frac{N_{\rm qp}^r}{V_{\rm tr}} \frac{\tau_{\rm qp}}{\tau_{\rm inj} - \tau_{\rm qp}} \left(e^{-t / \tau_{\rm inj}} - e^{- t / \tau_{\rm qp}}\right)\ .
\end{align}
We will use this pulse model in Section \ref{sec:resolutionandthreshold} to numerically model the energy threshold resulting from a pulse reconstructed from a simulated parity signal.

To close, we stress that the only fundamental physical limits in the above energy collection process are on $\eta_{\rm pb}$ in both the absorber and trap, and timescales associated with bulk anharmonic downconversion of phonons. All other efficiencies and time constants are in principle amenable to design optimization (for instance by limiting the amount of inactive absorbing material) or to material science study (surface-promoted phonon downconversion due to surface treatment). We summarize some of the quantities discussed in the previous sections in Table \ref{tab:eff} and provide baseline values we use in the rest of the paper.

\begin{table}[!h]
    \centering
    \begin{tabular}{|c|c|c|} \hline
      \textbf{Efficiency} & \textbf{Estimate} & \textbf{Notes} \\ \hline
      \textbf{$\eta_{\rm ph}$} (from Eq. \ref{eq:abs}) & 0.65 &  $f_{\rm surf}=0.04$ \\ 
      & & $N_{\rm surf}\approx10^3$ \cite{golwala2022novel} \\
      & & $f_{\rm abs}\approx0.1$ \cite{golwala2022novel} \\
      & & $f_{\rm loss}\approx0.35$ See Sec. \ref{sec:phonon} \\ \hline
      $\eta_{\rm pb,abs}$ \& $\eta_{\rm pb,trap}$ & 0.5 & Estimated from Refs. \cite{guruswamy2014quasiparticle, kozorezov2000quasiparticle}  \\ \hline
      $\eta_{\rm ce}$ & 0.32 & $\eta_{\rm ph}\eta_{\rm pb}$ \\ \hline
      $\eta_{\rm tr}$ & $0.4-0.7$ & Geometry dependent, \\ 
      & & see Appendices \ref{app:diffsketch}, \ref{app:depositresolution} \\ \hline
      $\Delta_{\rm abs}/\Delta_{\rm trap}$ & 4& For Hf/Al combo \\ \hline
    \end{tabular}
    \caption{Summary of efficiencies and estimated values, as described in Sec. \ref{sec:phonon} and \ref{sec:QPResponse}, and visualized in Fig. \ref{fig:effstring}.} 
    \label{tab:eff}
\end{table}

\subsection{QPD Energy and Charge Parity} \label{sec:qpd}

The QPD Hamiltonian is described by the modified CPB Hamiltonian \cite{martinis2004superconducting},
\begin{equation}
    \label{eq:cpb_hamiltonian}
    H_{\text{CPB}} = 4 E_C \left(\hat{n} - n_g + \frac{P-1}{4}\right)^2 - E_J \cos{\hat{\phi}},
\end{equation}
where we have included the charge parity-dependent shift, $P=\pm1$, to the offset charge $n_g$ \cite{serniak2019direct}. As is standard for a superconducting qubit, $\hat{n}$ and $\hat{\phi}$ are conjugate operators corresponding to the number of Cooper pairs on the island and the phase difference across the junction, respectively. $E_J=h\Delta/8R_Ne^2$ is the Josephson energy, the barrier tunneling energy of Cooper-pairs, with $R_N$ the normal-state resistance of the junction. $E_C=e^2/2C_{\Sigma}$ is the charging energy interpretable as the energy required to shift the island charge by an electron. $C_{\Sigma}$ is the total capacitance between the island and its circuit environment, formed from the sum of gate ($C_g$), junction ($C_J$), and shunt ($C_s$) capacitances (see Fig. \ref{fig:operation}). Following Ref.~\cite{serniak2019direct}, a charge parity switch event, from $P=+1$ (even) to $P=-1$ (odd), shifts the offset charge by half of a Cooper pair and hence shifts the energy spectrum. We label the eigenstates of the CPB Hamiltonian by the state index $i=0,1,2,...$ and charge parity terms $p=e,o$. 

\begin{figure}[!h]
    \includegraphics[width=0.48\textwidth]{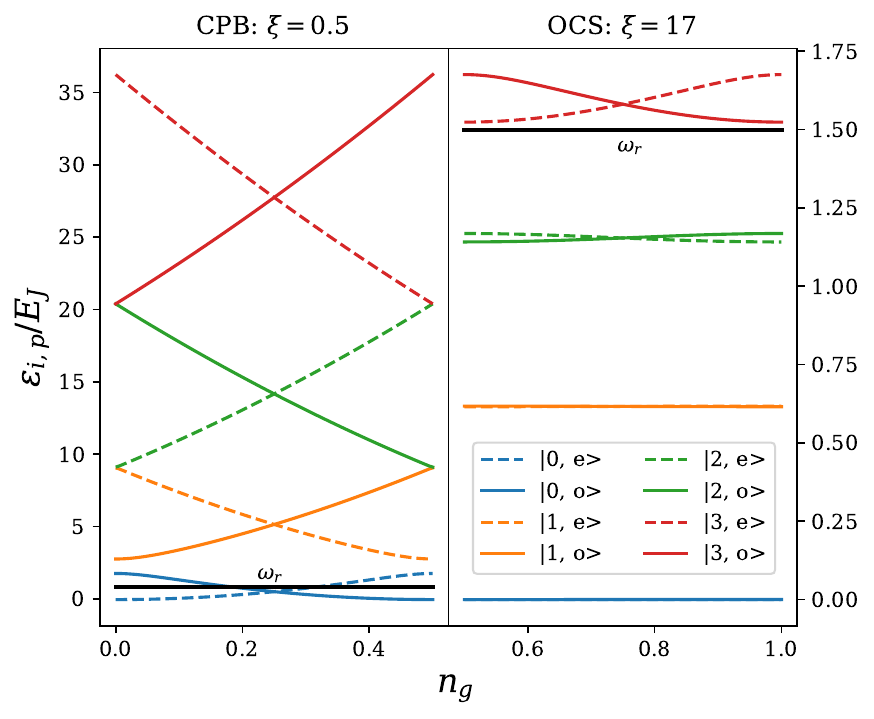}
    \caption{\label{fig:split} Example energy level structure from solving the CPB Hamiltonian for CPB (\textit{Left}) and OCS (\textit{Right}) $\xi$ ratios, with $E_J/2\pi\sim$5~GHz. The placement of the readout resonator is given by the black line.}
\end{figure}

We consider two limits of the CPB Hamiltonian, corresponding to both styles of QPD, which are distinguished by the dominant energy scale: $\xi=E_J/E_C$. For a CPB, the use of a small (few~fF) junction capacitance and the absence of shunt capacitance yield a large charging energy and $\xi \lesssim 1$. In the CPB limit, due to the curvature of the ground state, the difference between even and odd charge parity energy levels $\delta E$ can be large and of the same order as $E_J$. For an OCS, a large $\mathcal{O}$(100~fF) shunt capacitor, added to flatten the energy levels in order to decrease sensitivity to environmental charge~\cite{koch2007charge}, results in a smaller charging energy and $\xi \gtrsim 1$.  Note that only the lower energy levels are substantially flattened; the higher energy levels still exhibit a significant energy difference between even and odd states. As we operate the qubit in the ground state even in the OCS case, a frequency shift in the readout resonator is now the result of coupling to the even-odd energy difference of higher energy levels. Solutions to this Hamiltonian and respective even-odd energy spectra for a suggestive set of parameters are shown in Fig \ref{fig:split}. 

\subsection{Quasiparticle Tunneling} \label{sec:quasiparticle}

For QPD operation, two characteristic tunneling rates are relevant: the rate for quasiparticles to tunnel into the island or pad, $\gin$, and the rate for quasiparticles to tunnel out of the island or pad, $\gout$. From symmetry considerations, assuming equal absorber sizes, an OCS QPD has \gin=\gout. Closed-form expressions can be calculated for $\gin$ and $\gout$ assuming that the density of non-equilibrium quasiparticles, $\dnqp$, is uniform throughout the absorber \cite{lutchyn2005quasiparticle,lutchyn2007kinetics,catelani2011relaxation}. For typical device parameters, this approximation is justified since the quasiparticle diffusion timescale is much shorter than the quasiparticle tunneling or quasiparticle recombination timescales (see Appendix \ref{app:diffusion}). Under this assumption, one can define effective chemical potentials for the more general case of the CPB with absorber and island  \cite{shaw2008kinetics} of $\delta \mu_{\rm abs,island}$ (see Appendix \ref{app:closed} for specifics)

Using these effective chemical potentials, the rate for a thermalized quasiparticle to tunnel from the absorber to the island is given by Fermi's Golden Rule \cite{shaw2008kinetics,shaw2009qcd,lutchyn2007kinetics}:
\begin{eqnarray} \label{eq:gamma_in}
    \Gamma_{\text{in}} = && \frac{16 E_J}{h \Delta}  
    \int\displaylimits_{\Delta}^{\infty} dE \frac{E(E + \delta E) - \Delta^2}{\sqrt{(E^2 - \Delta^2)[(E + \delta E)^2 - \Delta^2]}} \nonumber \\ 
    && \times f(E - \delta \mu_{\rm abs}) [1 - f(E + \delta E - \delta \mu_{\rm island})],
\end{eqnarray}
where $f(\epsilon)$ is the Fermi-Dirac distribution. The energy difference between odd and even parity for the ground state, $\delta E$, renders the island a potential well of depth $\delta E$ (for the electron component of an absorber quasiparticle), effectively shifting the energy spectrum of the island states $\delta E$ below that of the absorber. Because $k_B T, \delta E \ll \Delta$, the Fermi level is well below $\Delta$ and the $1 - f(\epsilon)$ factor does not block tunneling: $\Gamma_{\rm in}$ is determined entirely by the absorber quasiparticle density, even in an OCS.  

It is important to note, however, that since we have $E_c \gg E_J$ in the CPB limit, we therefore  have $\delta E \gg E_J$ \cite{lutchyn2007kinetics}. Thus when a quasiparticle has tunneled on to the island, the island energy spectrum shifts upward by $\delta E$ \textit{and} the lowest energy state for a quasiparticle on the island is occupied, blocking the tunneling of additional absorber quasiparticles~\cite{averin1986coulomb} until the quasiparticle on the island tunnels out.

In an OCS, because $E_C \ll E_J $, there is no such blockade. Thus, any number of quasiparticles can tunnel back and forth between absorbers, and $\Gamma_{\rm in} = \Gamma_{\rm out}$.

For the OCS case, we can move away from the integral form for $\gin$ and work with the linearized form for small $\nqp$ (see Appendix~\ref{app:closed} for derivation):
\begin{equation} \label{eq:K_equation}
    \gin \approx \frac{16E_{J}k_{B}T}{\mathcal{N}\Delta h}n_{\rm qp}\equiv Kn_{\rm qp}.
\end{equation}
Because the above form neglects $\delta E$, it is only approximately correct for the CPB case. Regardless, $\gin$ is proportional to the quasiparticle density for both architectures. One can engineer the generic tunneling proportionality constant, $K$, by changing $E_J$, $\Delta$, $\mathcal{N}$ (the quasiparticle density of states defined in Appendix \ref{app:closed}), or $\delta E$ if applicable. Eq. \ref{eq:K_equation} is a key result of our work, providing for a closed form tunneling rate expression that can allow for both event reconstruction and detector optimization.

$\gout$ is more complicated for a CPB because of the aforementioned shift of the energy spectrum after tunneling has occurred.  The island quasiparticle now has the same energy as the gap energy of the absorber, the energy it had before tunneling \footnote{Because of the density of states singularity at $\Delta$, it is reasonable to assume all non-equilibrium quasiparticles in the absorber reside at the gap energy}.  Therefore, one must calculate the ``elastic'' tunneling rate, which is valid more generally when quasiparticles tunnel out after tunneling in without first relaxing to a lower energy. For a full discussion of the lead-to-island tunneling rate in the CPB case, see \cite{shaw2008kinetics}. In Sec. \ref{sec:resolutionandthreshold}, we outline how the differences in the tunneling rate of OCS and CPB devices affect device performance and expected energy thresholds.

\subsection{Readout Details} \label{sec:readout}

As is typical, we read out the charge parity state of the QPD by capacitively coupling it to a superconducting resonator~\cite{shaw2009qcd,serniak2019direct} in the dispersive limit \cite{krantz2019quantumengineer}. There is extensive experience with this approach in the literature for charge parity state detection, particularly for QCDs~\cite{echternach2018single,echternach2021large}. The use of a readout resonator decouples the qubit design ($E_J$ and $E_C$ particularly), from the requirements on the readout \textemdash\ the qubit can be optimized for energy sensitivity, and then the readout resonator adapted to the qubit design. Furthermore, the very low readout powers ($\lessapprox-140$~dBm) required for direct qubit-to-feedline coupling, along with the potential for unwanted, readout-power-generated quasiparticles seen in other superconducting systems \cite{de2012microwave}, makes a coupled resonator an attractive readout method. 

With a dispersive readout, the qubit is far detuned from the resonator, preventing direct exchange of energy between the two systems \cite{krantz2019quantumengineer}. Instead, the state of the qubit shifts the resonant frequency of the readout resonator from its bare value by an amount $\chi_{i,p}(n_g)$ (for qubit state $i$ and charge parity state $p$) which can be measured through a transmission measurement $S_{21}^{i,p}$. 

The tunneling proportionality $K$ from Sec. \ref{sec:quasiparticle} suggests that tunneling rates can approach for $\mathcal{O}$(100)~kHz for reasonable shifts in quasiparticle density. In order for the resonator to readout such a high tunnel rate with good fidelity, the readout resonator's signal bandwidth, $f_r/Q_r$, must be larger than the tunneling rate. A loaded quality factor of $\mathcal{O}(10^4)$ dominated by the coupling ($Q_r \approx Q_c$), in a $\mathcal{O}$(GHz) resonator, would meet said requirement by providing $\mathcal{O}$(MHz) signal bandwidth.

A key difference between CPB and OCS style QPDs is the mechanism responsible for the shift in resonator frequency. For CPBs, the existence of a strong relationship between gate charge and energy is interpretable as an effective quantum capacitance \cite{shaw2009qcd},
\begin{equation}
    \label{eq:quantum_capacitance}
    C_{i,p} = - \frac{C_g^2}{4e^2}\frac{\partial^2 \epsilon_{i,p}}{\partial n_g^2}.
\end{equation}
The change in this charge parity state-dependent capacitance $C_{i,p}$ shifts the readout resonator frequency \cite{shaw2009qcd}. The shift can be approximated in the reduced two-state Hilbert space case as,
\begin{equation} \label{eq:capacitance_shift}
    C_{i=(0,1),p} \approx 4\frac{C_g^2}{\xi C_\Sigma} \to 
    \chi_{i=(0,1),p} \approx -\frac{1}{2} \omega_r^3 L_r C_{i=(0,1),p}
\end{equation}

\begin{figure*}[!ht]
    \subfloat{%
        \includegraphics[width=0.49\textwidth]{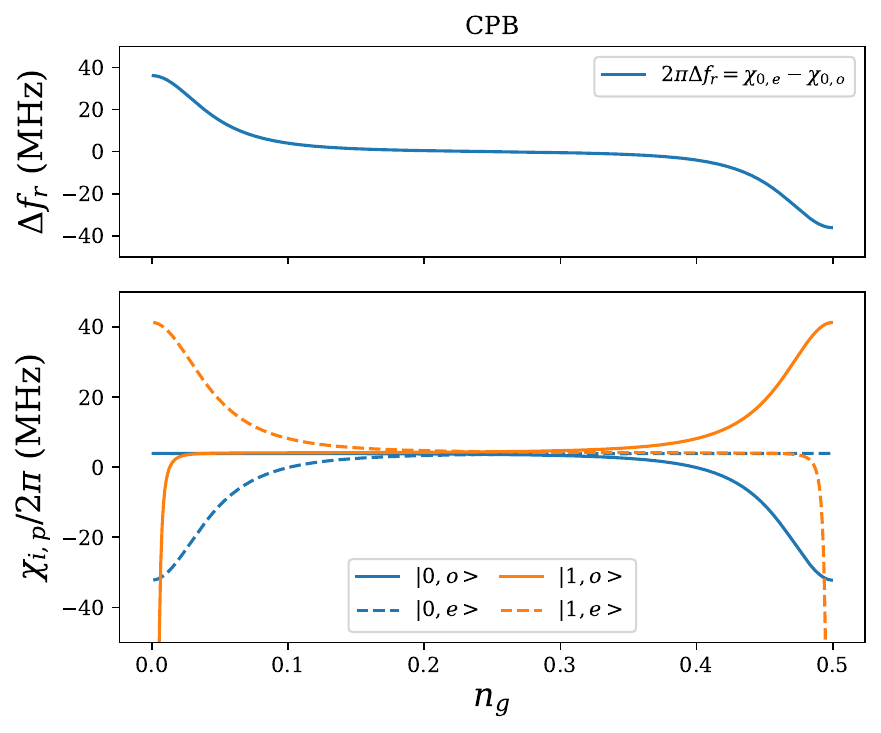}%
    }\hfill
    \subfloat{
       \includegraphics[width=0.49\textwidth]{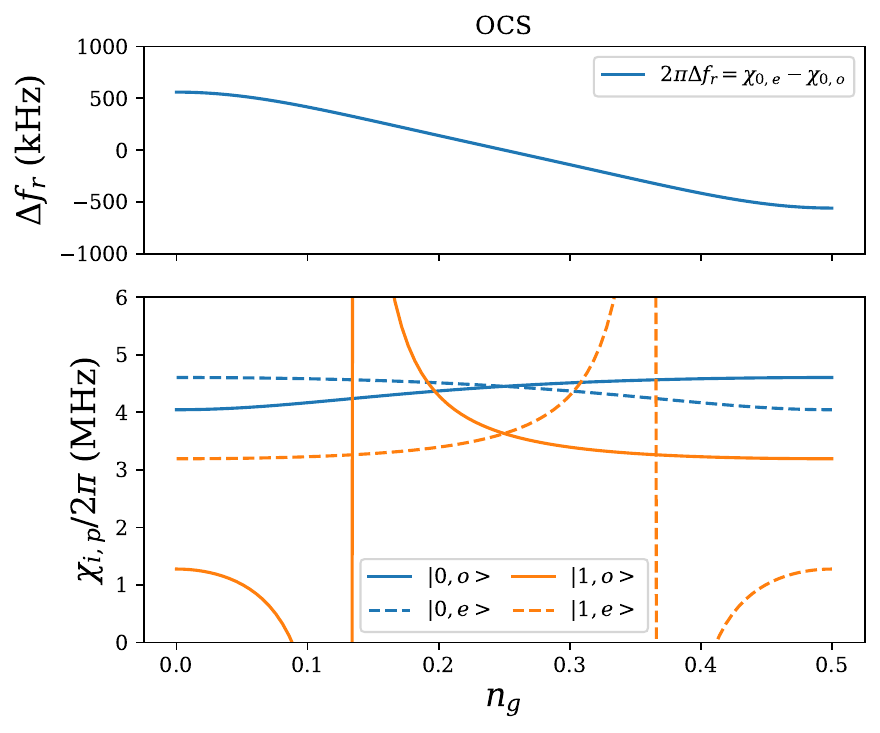}%
    }
    \caption{\label{fig:shifts} {\it Top:} Absolute difference in resonator frequency between even and odd charge parity for the ground states of a CPB (\textit{Left}) and OCS (\textit{Right}). {\it Bottom:} Shift in resonator frequency of the ground and first excited state of a CPB due to the quantum capacitance shift (\textit{Left}) and an OCS due to the dispersive shift (\textit{Right}).}
\end{figure*}

In the transmon regime ($\xi \gg 1$), Eq.~\ref{eq:capacitance_shift} implies the quantum capacitance vanishes. Instead, the shift in $\omega_r$ is dominated by an alternative shift \cite{serniak2019direct}. This $\chi_{i,p}$ can be derived from the full CPB-resonator Hamiltonian (see Ref.~\cite{manucharyan2012superinductance}) using second-order perturbation theory, 
\begin{equation}
    \label{eq:dispersive_shift}
    \chi_{i,p} = g^2 \sum_{j \neq i} \frac{2 \omega_{ij,p} |\bra{j, p} \hat{n} \ket{i, p}|^2}{\omega_{ij,p}^2 - \omega_r^2},
\end{equation}
where $\omega_{ij,p}$ is the transition frequency between transmon states $\ket{i,p}$ and $\ket{j,p}$, and $g$ is the coupling between the transmon and the resonator. As with the Lamb Shift in atomic orbitals, photons in the readout resonator, even if not resonant with the qubit transitions, can be virtually absorbed and reemitted by the qubit, causing a shift in the energy of $|i,p\rangle$. Because of the denominator in Eq.~\ref{eq:dispersive_shift}, the effect is larger the closer the photon is to a qubit transition, and so we engineer the largest possible shift by placing $\omega_r$ close to an energy level difference $\omega_{j,p} - \omega_{i,p}$. We choose to place $\omega_r$ close to $\epsilon_{03}$, as also shown in the right panel of Fig. \ref{fig:split}.

Assuming reasonable parameters for various components, we can see $\chi_{i,p}(n_g)$ for the CPB and OCS in Fig. \ref{fig:shifts}, along with the absolute difference in resonant frequency $\Delta \omega_r^{i} = (\chi_{i,e} - \chi_{i,o})$. We see that for ground state shifts, $\Delta \omega_r^{\rm max}/2\pi=\mathcal{O}({\rm MHz})$ for the CPB and $\Delta \omega_r^{\rm max}/2\pi=\mathcal{O}(100)$~kHz for the OCS. The CPB shift is of the same size as the FWHM and is eminently resolvable. While the OCS shift is smaller, the nominal frequency uncertainty of a thin-film superconducting resonator, read out at low power and limited by $~4K$ amplifier noise, is $\mathcal{O}$(100) Hz. \cite{burnett2018noise,ramanathan2024significant}. These shifts should be easily resolvable at expected operating powers ($\sim$-90~dBm at device) without special attention to readout strategy and crucially without requiring the use of any quantum-limited amplifier with sub-Kelvin noise temperatures.

\begin{figure}[!t]
    \includegraphics[width=0.48\textwidth]{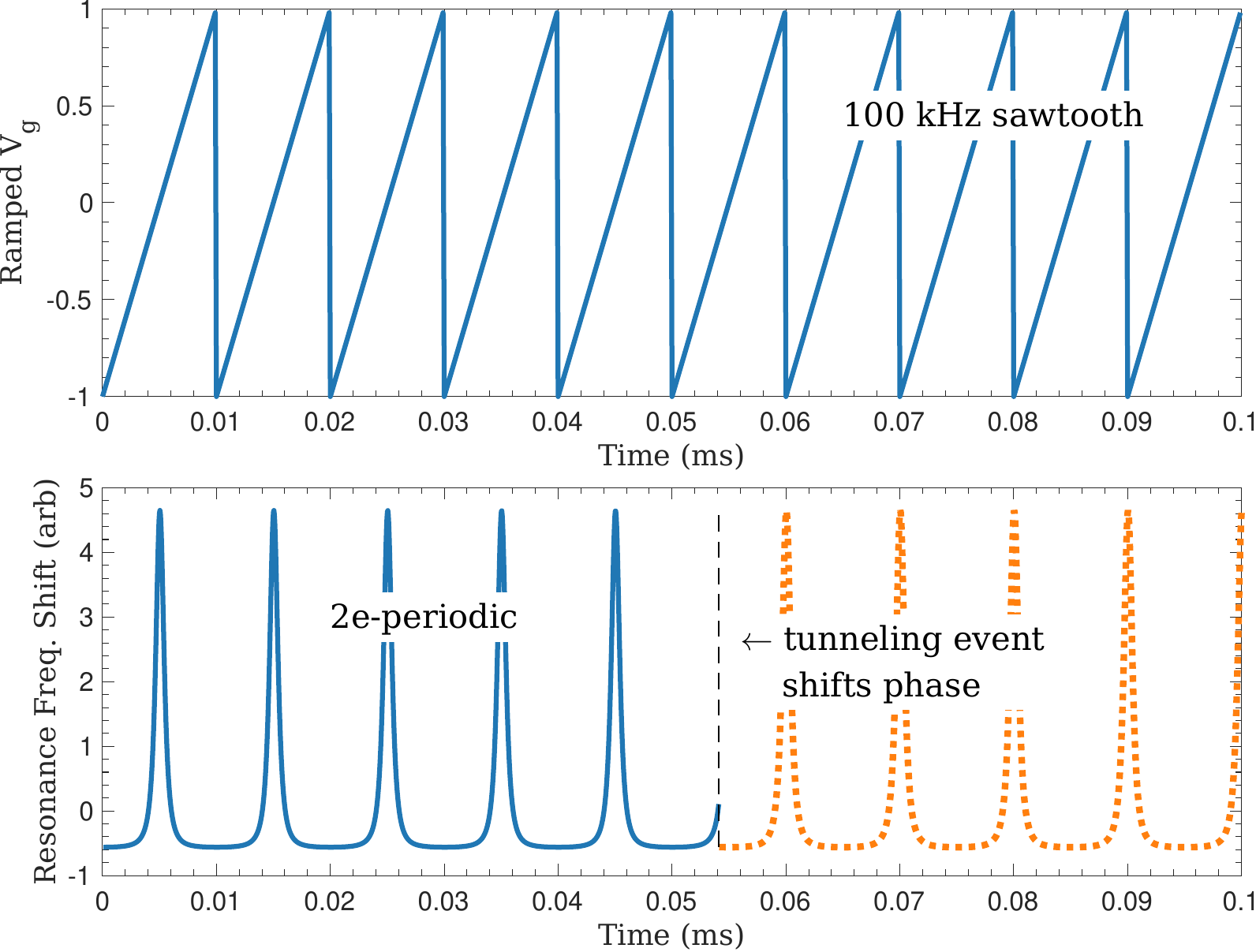}
    \caption{\label{fig:ramp} {\it Top:} Example of a sawtooth voltage bias sweep at 100 kHz. {\it Bottom:} The stylized frequency shift expected when in the even state is given by the blue curve. A tunneling event in an OCS QPD effectively shifts the phase of this signal by shifting from the even$\to$odd curves and introducing a noticeable gap in the output signal.}
\end{figure}

One complication in operating multiple QPDs on the same feedline (for use cases requiring pixelization) is that a gate bias value $V_g$ will not match the optimal operating point for all sensors, due to inherent uncorrelated static charge drifts for each sensor. As Fig. \ref{fig:shifts} makes clear, one needs to operate at half-integer $n_g$ values corresponding to peaks in the resonator response. A potential solution is to \textit{sweep} the bias in a similar fashion to Ref. \cite{echternach2018single}. Briefly, by ramping the bias in a sawtooth manner at a frequency $f_{\rm ramp}$ as shown in Fig. \ref{fig:ramp} Top, one traces through an entire cycle of $n_g$ for all sensors. In an OCS QPD, tunneling events will break the smooth variation in $\chi_{i,p}$ and this effective phase shift is now the signature of a single tunneling event. The caveat is that if the period of the ramp signal is significantly longer than the expected signal lifetime ($\sim$3$\tau_{\rm qp}$), a quasiparticle burst may occur as the sweep passes the point where $\delta E = 0$ between the even and odd states, hence a large portion of the signal may be lost as $\Delta f_r$ dips below the noise floor. This issue can be partially addressed by ensuring that the period of the ramp signal is less than half of the expected signal lifetime; in this case you will only lose the fraction of signal roughly equal to the fraction of the periodic ramp where $\Delta f_r$ dips below the noise floor (<10\% with conservative estimates). This technique is more complicated for CPB QPDs, as the relative energy level of a quasiparticle on the island is higher for $n_g=(0,2e,...)$ \cite{echternach2013photon} than in the absorber as the island depth $\delta E$ flips sign with the changing $n_g$ as seen in Fig. \ref{fig:stylized}. At and near these points then, the small CPB island is effectively instantaneously cleared. This behavior does not manifest for an OCS device as $\delta E$ is negligible with increasing $\xi$. Thus, in a CPB QPD, such a rapid gate sweep will preferentially keep the device in the even state. However, episodes of rapid tunneling, such as after a burst of produced quasiparticles, will suppress peaks in the trace and produce ``gaps" of one to a few peak widths in $\chi_{i,p}$ \cite{echternach2018single} before the periodic trace resumes. While the number of tunneling events cannot be ascertained from the presence of a gap, it can be used to tag discrete energy deposits that generated the rapid tunneling \textemdash\ potentially from single phonon absorption.

\section{Device Performance} \label{sec:deviceperformance}

Computing QPD performance, through modeling of expected signal and background rates, requires grappling with a variety of challenges \textemdash\ including the stochastic nature of the discrete charge parity transitions and asymmetric tunneling (at least in the case of the CPB) among other issues. To begin our discussion, we catalog expected fundamental and readout related noise sources in Sec. \ref{sec:noisesources}. Next, we investigate a simplified device model in Sec. \ref{sec:estimates}, considering only OCS QPDs to get a flavor of expected performance metrics. Here we express our results in the more analytically tractable unit of number of transitions within a given period. Finally, Sec. \ref{sec:resolutionandthreshold} provides a more robust numerical modeling, using charge parity state dwell-times instead and taking into account the effects of pulsing and asymmetric time-varying tunneling rates, and further discusses the factors that may affect our ability to disentangle signals from background. 

For concreteness, we provide parameters for four hypothetical devices in Table \ref{tab:devices}, investigating different trap material choices (aluminum\footnote{assuming that the absorber in this case is an unspecified higher gap material like $\alpha$-Ta or Nb} vs hafnium) and device styles. The Hf case reflects an optimistic, yet plausible, scenario regarding the use of a new material with favorable properties and thermal backgrounds. To avoid the confounding effects of backgrounds (e.g. radiogenic, cosmogenic, or other otherwise unexplained excesses \cite{baxter2025low}), the choice of substrate, geometry-dependent phonon-to-film couplings, and quasiparticle diffusion, we only deal with absorbed energy \Eabs\ bounds, which roughly reflects intrinsic sensor properties.

\begin{table}[!h]
    \centering
    \begin{tabular}{|c|c|c|c|c|} \hline
      & \multicolumn{2}{c|}{\textbf{Aluminum}} & \multicolumn{2}{c|}{\textbf{Hafnium}} \\ \hline
      Style & CPB & OCS & CPB & OCS \\ \hline
      $V_{\rm tr}$ & 100 \umthree & 50 \umthree ($\times2$) & 1000 \umthree & 500 \umthree ($\times2$) \\ \hline
      $T_{\rm c}$ & \multicolumn{2}{c|}{1.2~K} & \multicolumn{2}{c|}{0.25~K} \\ \hline
      $T_{\rm opr}$ & \multicolumn{2}{c|}{0.1~K} & \multicolumn{2}{c|}{0.025~K} \\ \hline
      $\Delta$ & \multicolumn{2}{c|}{190~$\mu$eV} & \multicolumn{2}{c|}{40~$\mu$eV} \\ \hline
      \Eabs & \multicolumn{2}{c|}{200 meV} & \multicolumn{2}{c|}{40 meV} \\ \hline
      \nought & \multicolumn{2}{c|}{0.3 \umminusthree} & \multicolumn{2}{c|}{0.03 \umminusthree$^{**}$} \\ \hline
      $E_J / h$ & 4.9 GHz & 6.14 GHz & 4.9 GHz & 6.14 GHz \\ \hline
      $E_C / h$ & 11.1 GHz & 356 MHz & 11.1 GHz & 356 MHz \\ \hline
      $\xi$ & 0.5 & 17 & 0.5 & 17 \\ \hline
      Fano factor $F$ & \multicolumn{4}{c|}{0.2} \\ \hline
      \tqp & \multicolumn{4}{c|}{1~ms$^\dagger$} \\ \hline
      $\tau_{\rm inj}$ & \multicolumn{4}{c|}{2~ms} \\ \hline
      $K$ & \multicolumn{2}{c|}{3 kHz$\cdot$\umthree} & \multicolumn{2}{c|}{20~kHz$\cdot$\umthree} \\ \hline
      $\Gamma_{\rm out}$ & 2 kHz & n/a & 10 kHz & n/a \\  \hline
    \end{tabular}
    \caption{Basic operating characteristics for device examples discussed in text with example \Eabs. $^\dagger$Literature values of Hf thin-film quasiparticle lifetimes are $\sim$400 $\mu$s \cite{zobrist2019design}, though to our knowledge, and unlike Al, no serious R\&D effort has been undertaken to extend this \textemdash which suggests significant improvement can be achieved in this frontier. $^{**}$Assuming thermal quasiparticles comprise the entirety of the quiescent background.} 
    \label{tab:devices}
\end{table}

\subsection{Noise Sources} \label{sec:noisesources}
To properly estimate the performance of a QPD, we list some expected contributions to the noise and how they will be modeled in the sections that follow.

\textbf{Telegraph Noise:}  A pulse in the quasiparticle tunneling-in rate results in a two-state time-dependent charge parity response in the QPD. This two-state flip-flop is referred to as a \textit{telegraph} process. This telegraph noise represents noise \textit{during} a pulse that may affect energy resolution at non-zero energy. For determining the threshold from the baseline (zero-energy) noise, we focus on the shot noise on the quiescent tunneling rates $\Gamma_{\rm in,out}$, where the variance on the number of tunneling events in a given time period equals the expected mean number of tunneling events in that period.

\textbf{Residual Quasiparticle Noise:}  For thin film aluminum, a quiescent (residual or steady-state) quasiparticle density of $n_0=0.01$\textendash1$~\mu$m$^{-3}$ has been measured in qubit systems \cite{connolly2023coexistence}, with the population influenced by thermal or other pair-breaking processes \cite{de2012generation}. We can account for the effect of this population by again modeling it as a super-Poissonian shot noise with variance proportional to the total quasiparticle number $\sigma^2 = 16n_0V$ (the prefactor comes from including pairing effects for phonon to quasiparticle breaking and recombination rates).

\textbf{Fano Noise: } For a given phonon energy deposit, the same number of quasiparticles will not always be generated. The process is not, however, Poissonian because of correlations between the creation process for the quasiparticles. The sub-Poissonian nature is encoded in the Fano factor~\cite{fano1947ionization}. We include this contribution in our energy resolution modeling with $\sigma^2 = F \Nqpr$. As a first pass, we take $F$ to be $\approx 0.2$~\cite{verhoeve1997superconducting}, a number calculated for quasiparticle production from photon absorption $\gg 2\Delta$. Its application to low energy phonon absorption will need to be experimentally validated. 

\textbf{System \& Amplifier Noise:} 
Amplifier noise can potentially limit the ability to observe charge parity transitions.  The fractional noise on resonance due to amplifier noise is given by \cite{wen2022performance},
\begin{equation}
    \frac{\delta f}{f_r} \approx \sqrt{\frac{k_BT_N\Gamma_{\rm max}}{2P_g}}\frac{Q_c}{Q_r^2} 
\end{equation}
For conservative assumptions of noise temperature $T_N = 10$~K, feedline (readout) power of $P_g=-120$~dBm, coupling quality factor $Q_c = 10^4$ ($\approx$ the loaded value $Q_r$), and a resonator bandwidth set in concordance with the maximum expected tunneling rate $\Gamma_{\rm max}$ of 1~MHz (corresponding to the expected transition rate given a 10 eV energy deposition directly into an OCS device absorber), the fractional frequency noise is 70~kHz. This is $\ll \Delta f_r$ from Sec. \ref{sec:readout} (even for the smaller shifts seen by the OCS style QPD). 

\textbf{TLS Noise:} Two-level system (TLS) noise is often significant for thin-film superconducting resonators \cite{zmuidzinas2012superconducting, ramanathan2024significant}. TLS has also been seen to limit qubit coherence lifetimes \cite{abdurakhimov2022identification}. We can calculate the impact of TLS noise in the same way as for amplifier noise: we integrate a typical TLS noise PSD~\cite{noroozian2009two,zmuidzinas2012superconducting}, $S_{df/f}$, over a similar bandwidth $\Gamma_{\rm max}$ as for amplifier noise and find,
\begin{equation}
    f_r\sqrt{\mathcal{S}_{\rm \delta f/f}\Gamma_{\rm max}}\sim10\,{\rm kHz},
\end{equation} 
As in the case of amplifier noise, the RMS noise is much smaller than the charge parity transition signal and so does not factor into sensitivity.

\subsection{Performance Estimates} \label{sec:estimates}

We are now in a position to provide estimates for QPD performance. For a given time window $t_w=t_f-t_0$, the expected number of transitions $S_w$, can be computed as an inhomogeneous Poisson process by integrating the expected pulse shape,
\begin{equation} \label{equation:swabstract}
    S_w = \int_{t_0}^{t_f} K \dnqp(t) dt.
\end{equation}

For our pulse model from Eq. \ref{eq:qp_pulse}, the total expected number of transitions from the entire pulse (with $t_w\gg\tau_{\rm inj,qp}$) is,
\begin{equation} \label{eq:swclosed}
     S_{\rm avg} = \frac{K}{V_{\rm tr}} \Nqpr \tau_{\rm qp}.
\end{equation}
We note the straightforward dependence on increasing the signal transitions, by increasing the tunneling proportionality (for instance by increasing $E_J$), decreasing the trap volume $V_{\rm tr}$ and increasing $\tau_{\rm qp}$. 

A simple expression for the stochastic background can be obtained by defining the random variables:
\begin{align} \label{eq:backgroundflips}
    \boldsymbol{N_{0}} &\sim{\rm Gaussian}(N_{0},\sigma=4\sqrt{N_{0}}) \nonumber \\
    \boldsymbol{B}     &\sim{\rm Poisson}(B_{\rm avg}=\frac{Kt_w}{V_{\rm tr}}\boldsymbol{N_{0}}),
\end{align}
with $N_0$ the quiescent quasiparticle population (in quasiparticle number units) and $\boldsymbol{B}$ is the background number of transitions within the window, taking into account the telegraph shot noise. Note that $\boldsymbol{B}$ is a mixture distribution. In the limit of a large number of transitions we can treat $\boldsymbol{B}$ as a Gaussian process, which allows us to write the variance in the background distribution as the distributed product,
\begin{align} \label{eq:sigB}
    \sigma_{B_{\rm tot}}^2 &= \sigma_{N_0}^2 \left(\frac{dB}{dN_0}\right)^2 + \sigma_B^2 \nonumber \\ 
               &= \left(Kt_w+16\frac{K^{2}t_w^2}{V_{\rm tr}}\right)\nought.
\end{align}

The signal variance can be written in a similar way, including the telegraph and Fano components, as: 
\begin{align} \label{eq:nf}
    \boldsymbol{N_F} &\sim{\rm Gaussian(\Nqpr,\sigma_{F}=\sqrt{\it{F}\Nqpr})} \nonumber \\
    \boldsymbol{S}     &\sim{\rm Poisson}(\frac{K \tau_{\rm  qp}}{V_{\rm tr}}\boldsymbol{N_F}).
\end{align}
As we are interested in a sensitivity estimate here, we elide the changing transition rate due to the varying pulse shape and use only the overall number of transitions from Eq. \ref{eq:swclosed}. Treating $\boldsymbol{S}$ as Gaussian for tractability we have,
\begin{equation} \label{eq:sigS}
    \sigma_{S}^{2} = \left(K\frac{\tau_{\rm qp}}{V_{\rm tr}}+K^{2}\left(\frac{\tau_{\rm qp}}{V_{\rm tr}}\right)^{2}F\right)\Nqpr.
\end{equation}

Inputting the parameters from Table \ref{tab:devices} into equations \ref{eq:swclosed}, \ref{eq:backgroundflips}, \ref{eq:sigB}, \ref{eq:sigS}, we arrive at at the values in Table \ref{tab:flips}. 
\begin{table}[!h]
    \centering
    \begin{tabular}{c|c c} 
    & Aluminum & Hafnium \\ \hline 
    $S_{\rm avg}$ & 20 & 13 \\
    $B_{\rm avg} \equiv \langle \boldsymbol{B} \rangle $ & 3.6 & 2.4 \\
    $\sigma_{B_{\rm tot}}$ & 2.0 & 1.6  \\
    $\sigma_S$ & 8.9 & 7.1 \\ \hline
    \Eabs\ threshold (meV) & 165 & 37 \\ 
    \end{tabular}
    \caption{Expected signal \& background transition counts and resolvable absorbed energy threshold for OCS style devices, using the example parameters provided in Table \ref{tab:devices}, for a window $t_w=3\tau_{\rm abs}$.} 
    \label{tab:flips}
\end{table}
We have used Eq. \ref{eq:swclosed} to convert $\sigma_B$ into a ``5$\sigma$" threshold, the minimum energy deposit potentially resolvable from background. The values, particularly for the larger Hf device and extrapolating downward in volume, reflect the potential to resolve single phonon events.

The crucial point, we stress, is that it is the volume of the trap and \textit{not} the absorber that sets the tunneling rate. Referring to Appendix \ref{app:diffsketch}, we note that the fraction of quasiparticles tunneling into the trap $\eta_{\rm tr}$ can be engineered to be $\approx0.2$. Such a regime yields
$N_{\rm trap} \approx \mathcal{O}(0.1-1) \cdot N_{\rm abs}$, as the quasiparticle multiplication effects offset the trapping fraction. Thus, the number of quasiparticles produced in the much smaller Hf trap can be similar to that within the Al absorber.

Care needs to be taken to account for the quiescent quasiparticle density in the absorber and its effect on the quasiparticle density in the trap. Non-thermal mobile background quasiparticles created in the absorber, such as through a constant source of background IR radiation \cite{chang2023supercdms}, may impact the density in the trap because the trap acts as a sink. We can estimate the size of this effect using a simple balance model (see Appendix \ref{app:detailed}) that factors in generation, recombination, and trapping. For a seemingly negligible excess density of $n_0=0.05$~\umminusthree\ in reasonably sized Al absorber pads of $\sim$60000~\umthree\ total (2 $\times$ 400~$\mu m$ $\times$ 200~$\mu m$ $\times$ 600~nm), a 500~\umthree\ Hf trap would have a much higher (than purely thermal origin) quasiparticle density of $n_0^{\rm trap}\sim0.15$~\umminusthree. By propagating expected trapping and phonon efficiencies (as detailed in Appendix \ref{app:depositresolution}), we find for this system a resolution and threshold of energy \textit{deposited in the substrate} of $\sigma_{\rm dep} \approx 100\,\rm{meV}$ (i.e. $E_{\rm dep}^{\rm threshold} \approx 500\,\rm{meV}$).

It will be crucial then to drive down $n_0$ in the absorber to the thermal floor. In such a condition, the dominant background will again be the thermal density within the trap alone. Appendix \ref{app:depositresolution} computes the resolution on deposited energy for a 1~cm$^2$ crystal and 50 QPDs (for 4\% area coverage), resulting in 
\begin{equation} 
    \sigma_{\rm dep} \approx 70 \left(\frac{V_{\rm trap}}{800\,\umthree}\right)^{\frac{1}{2}}\ {\rm meV}.
\end{equation}

Such a design would immediately allow for the sensing of single optical phonons. Optimization of the trap volume down to 10s of \umthree\ with corresponding novel developments in quasiparticle trapping into new junction materials (e.g. hafnium) would result in a threshold of just a few 10s of meV, with the ability to probe for signals from the bulk distribution of acoustic phonons. All told, these calculations demonstrate the potential to \textit{detect} and \textit{reconstruct} $\ll$~eV substrate energy deposits, providing a path to sensitivities beyond currently demonstrated technologies~\cite{wen2022performance,tesseract2023results}.

\subsection{Energy Reconstruction \& Thresholds} \label{sec:resolutionandthreshold}

Having covered a basic model of the signal and noise, we are ready to work through more concrete, numerically simulated performance metrics. We simulate the two-state, charge parity signal of a QPD in response to an energy deposit, folding in the fundamental noise sources discussed in the previous section. We consider the case of a device consisting of a single superconducting material and discuss sensitivity on \Eabs, restricting ourselves once again to only the intrinsic sensor physics. These results can heuristically be extended to \Edep\ by factoring in the various collection efficiencies (i.e. $\eta_{\rm ce}$ and $\eta_{\rm pb}$) by hand.

\begin{figure}[!h]
    \includegraphics[width=0.48\textwidth]{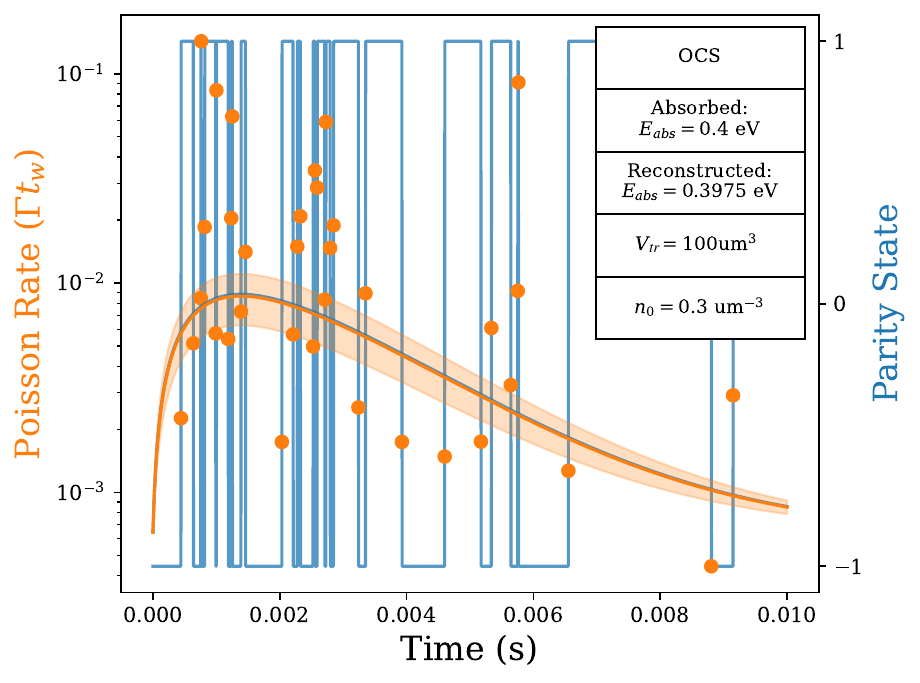}
    \caption{\label{fig:curvefit} Simulated charge parity state time-series (blue telegraph signal) for parameters given in inset. Dwell times points derived from this trace, converted to an equivalent transition rate, are given as the orange circles. The pulse reconstruction is shown by the orange line, with the shaded band corresponding to 1$\sigma$ on the reconstructed energy.}
\end{figure}

We start by discretizing pulses (defined by Eq. \ref{eq:qp_pulse}) and their time-spans into bins of width $t_w$, using values from Table \ref{tab:devices} in their construction. Next, for each time bin, we sample a Poisson distribution to determine whether charge parity state transitions occur in the given interval. In any experimental realization, the sampling frequency must also be chosen to ensure that the probability of more than one transition occurring in a window is negligible. The relevant Poisson parameter depends on both the current charge parity state, and whether the device is a CPB or OCS:
\begin{align*}
    \lambda_{\text{even}} &= S_w({\bf N_F}) + B_{\text{avg}} \\
    \lambda_{\text{odd}} &= \begin{cases}
        \lambda_\text{even} & \text{OCS} \\
        \Gamma_{\text{out}} t_w = \text{constant} & \text{CPB}
    \end{cases} 
\end{align*}
where $S_w({\bf N_F})$ should be interpreted as initially sampling ${\bf N_F}$ from Eq.~\ref{eq:nf}, taking that number to be \Nqpr\ in Eq.~\ref{eq:qp_pulse}, and subsequently computing $S_w$ from Eq.~\ref{equation:swabstract}. From this process, we can construct a telegraph signal time-series of transitions, an example of which is shown by the blue curve in Fig. \ref{fig:curvefit}. 

\begin{figure*}[!ht]
    \subfloat{
        \includegraphics[width=0.49\textwidth]{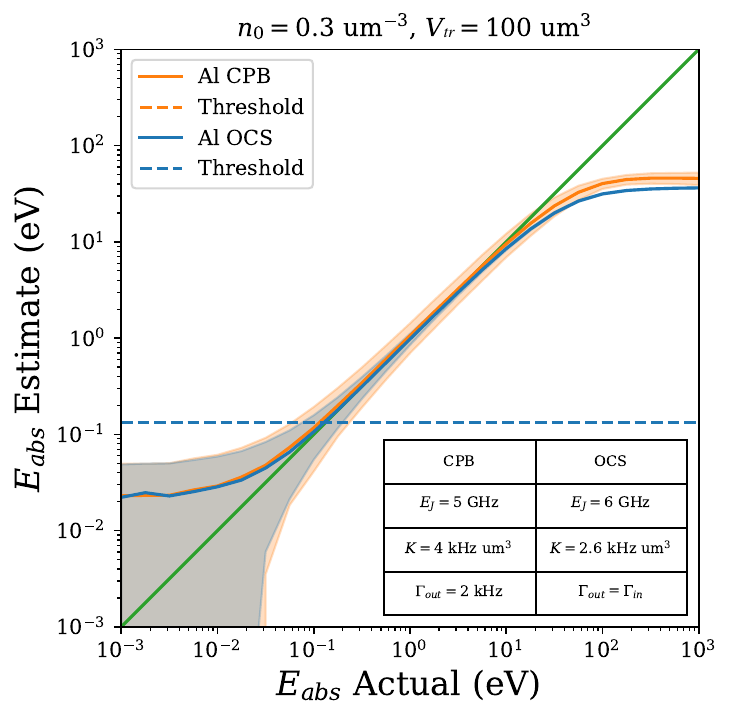}
    }\hfill
    \subfloat{
        \includegraphics[width=0.49\textwidth]{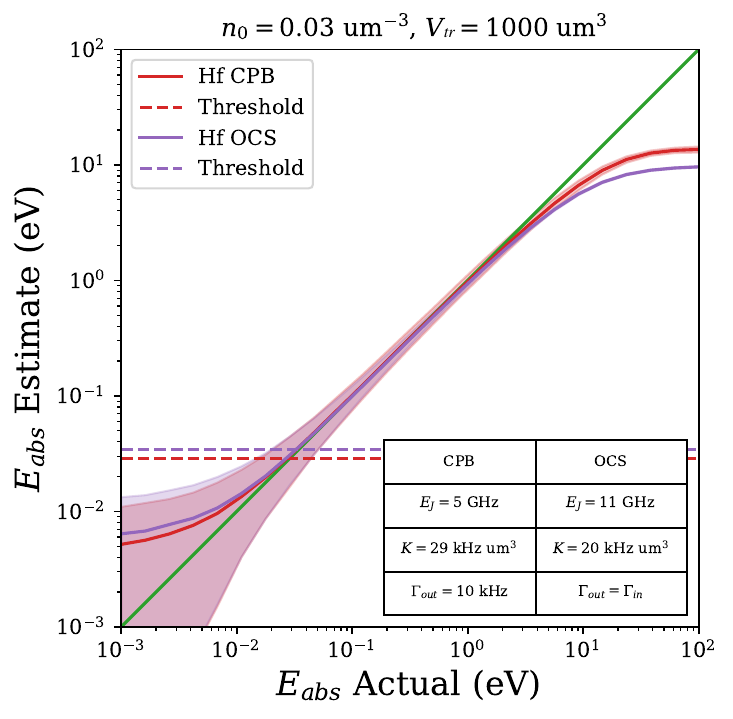}
    }
    \caption{\label{fig:E_abs_sweep} {\it Left:} Reconstructed energy estimate vs simulated energy absorbed by single Aluminum CPB \& OCS sensors with 100\umthree\ trap volume. {\it Right:} Similar curves Hf devices with 1000\umthree\ total volume of absorber. $1\sigma$ equivalent quantiles are given by the shaded bands, with energy thresholds (described in text) shown as the dotted lines.}
\end{figure*}

\begin{figure*}[!ht]
    \subfloat{
        \includegraphics[width=0.49\textwidth]{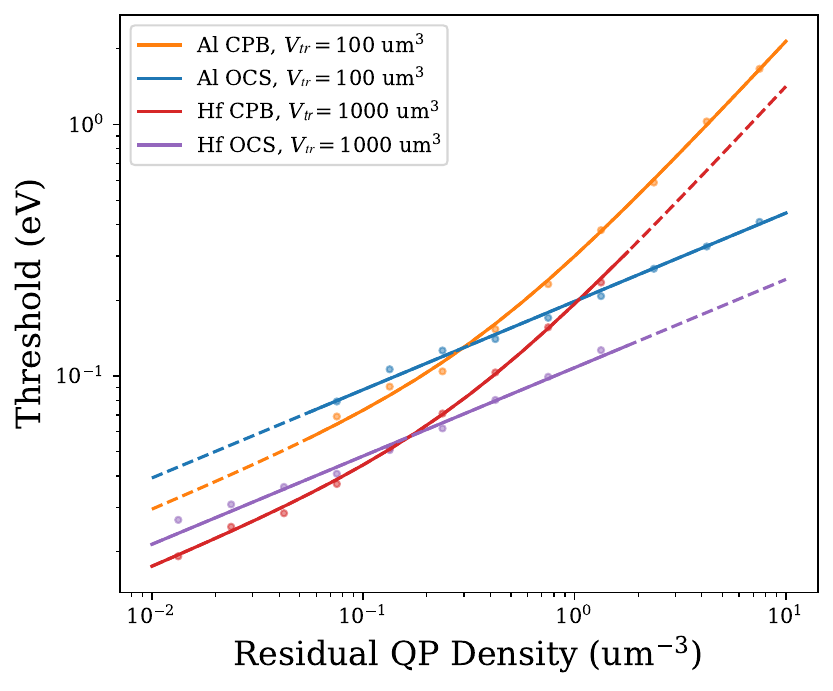}
    }\hfill
    \subfloat{
        \includegraphics[width=0.49\textwidth]{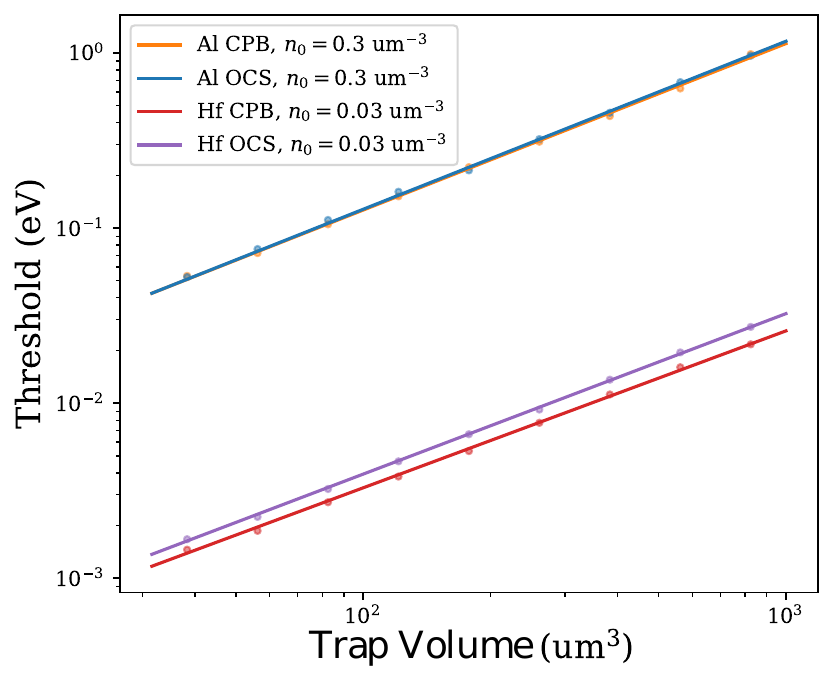}
    }
    \caption{\label{fig:paramsweeps} \textit{Left: } Simulated effect of quiescent quasiparticle density on \Eabs\ threshold. $n_0$ is simulated in the neighborhood of Table \ref{tab:devices} values and extrapolated based on expected behavior elsewhere. \textit{Right: } Simulated effect of volume on \Eabs\ threshold.}
\end{figure*}

To reconstruct a pulse, we first convert its telegraph signal to a series of even and odd state dwell times, $\delta t_{e,o}$, defined as how long the system stayed at its last charge parity until a transition occurred. We then estimate the Poisson parameter at each bin a transition occurs, as:
\begin{equation*}
    \hat{\lambda}_{\text{even, odd}} = \frac{t_w}{\delta t_{\text{e, o}}}
\end{equation*}
This prescription provides us with a series of points in time that are samples of the true transition rate, seen as the orange points in Fig. \ref{fig:curvefit}. We can fit this data with a 2 parameter model (\dnqp, $n_0$) incorporating the expected pulse shape and quiescent quasiparticle density, with results shown as the orange best-fit line in the figure. For this simulation, we assume we know the pulse start time, while a more complete simulation would fit it as a free parameter (and accept the corresponding impact on threshold).

There are a few subtleties that depend on the value of $\xi$. Because tunneling in is blocked when the CPB is in the odd state, the total number of observed qubit state transitions in any interval will be a factor of roughly $2\Gamma_{\rm out}/(\Gamma_{\rm in} + \Gamma_{\rm out})$ fewer in the CPB case (for equivalent $\Gamma_{\rm in}$). Furthermore, since the OCS tunneling is bi-directional, both even and odd dwell times provide information about the quasiparticle density whereas, in the CPB only, $\Gamma_{\rm in}$ is proportional to quasiparticle density.

For each of the device parameter cases provided in Table \ref{tab:devices}, we repeat this simulation thousands of times to characterize the energy threshold and reconstruction, shown in Fig. \ref{fig:E_abs_sweep}. The shaded bands reflect the $1\sigma$ deviation in reconstruction. For regions where the distribution of reconstructed events is non-normal, such as at zero to low energies, we instead compute $1\sigma$-equivalent quantiles. We infer ``$5\sigma$" thresholds from the baseline resolution reconstructed from simulations with no energy deposit. We note the consistency between these thresholds and the approximate values from Sec. \ref{sec:estimates}. For sufficiently low \Eabs, the residual quasiparticle density $n_0$ sets the threshold. At high \Eabs, saturation occurs in both sensor modalities when the tunneling rate exceeds the resonator bandwidth, $\gin > Q_r/\tau_{r}\pi$, and one cannot accurately reconstruct the number of transitions due to the slow resonator response. This result implies that the resonator should be designed to accommodate the typical tunneling rate, perhaps at a cost in $Q_r$ and thus in signal-to-noise for the charge parity transition relative to amplifier noise. 

While the OCS and CPB schemes have comparable thresholds (almost identical for the Al device due to judiciously chosen parameters), they differ in other performance metrics. The CPB suffers larger reconstruction variance (note the size of the bands in Fig. \ref{fig:E_abs_sweep}) because the constant nature of $\Gamma_{out}$ results in fewer transitions for a given energy deposit. Also, the CPB threshold degrades more dramatically with increasing $n_0$ than the OCS due to blocking of tunneling in. The two architectures' performance is only similar when $\Gamma_{in} \ll \Gamma_{out}$. It should be noted that, at high enough $n_0$, both modalities experience a degradation in sensitivity due to saturation, with the OCS saturating faster because of the higher number of transition per interval. The overall dependence on volume is as expected: a smaller volume provides a higher signal quasiparticle density for a given energy deposit, though this calculation neglects any potential related reduction in phonon collection efficiency. Also as expected, Hf offers lower threshold at the same $n_0$ than Al because of the higher quasiparticle yield per unit energy deposit. However, an increase in $n_0$ with decreasing $\Delta$ could counter this gain. Both QPD schemes have attractive features and deserve exploration for phonon sensing.

\subsection{Additional Considerations} \label{sec:additional}

In the preceding sections, we demonstrated that, for a reasonable set of assumptions and parameters, the QPD offers potentially ground-breaking energy threshold for phonon-mediated particle detection. A development program would begin with single-material devices to establish basic experimental parameters followed likely by two-material devices incorporating quasiparticle trapping.  Many choices must be made and challenges overcome, however, which we delineate here:

\textbf{Material Selection: } There are a plethora of low $T_C$ ($<1$~K) superconducting materials, including alloys, that are suitable for thin film applications \textemdash\ including W, AlMn, Ir, IrPt, Os, and Ti~\cite{saab2000design,mazin2022superconducting}. Workable tunnel junctions have not been convincingly demonstrated with most of these materials. Quasiparticle lifetime, phonon acoustic mismatch and collection, and amenability to device fabrication not been exhaustively investigated. R\&D on these and other lower-gap materials is warranted given the potential gain in responsivity. 

\textbf{Volume \& Diffusion: } We have already discussed the effects of absorber/trap sizing and volume on tunneling rates in Sec. \ref{sec:deviceperformance}. However, the discussion there uses the shorthand of chemical potentials \textemdash\ convincingly used in the original QCD detector papers to demonstrate the linearity of the tunneling rate \textemdash\ while a more complete modeling would take into account the dynamic quasiparticle drift and diffusion processes. Those topics are, however, fraught with uncertainty, as briefly discussed in Appendix \ref{app:diffusion}, with varying literature reported values for diffusion coefficients and its dependence on temperature, film thickness, and gap \cite{yen2015phonon,yen2016quasiparticle,hsieh1968diffusion,dong2022measurement,fink2022gram}. Experimental verification, through direct measurement of QPD response, will prove crucial for determining future optimizations of absorber shape and sizing.

\textbf{Quasiparticle Cascade and Trap Design: } As exhaustively studied in the CDMS dark matter direct-detection experimental program (e.g. see Refs. \cite{saab2002search, pyle2012optimizing, kurinsky2018low} for theses spanning two decades), efficiently collecting quasiparticles within a constrained region, in absorber$\to$trap designs, will be of paramount importance in maximizing phonon sensitivity. As is partially discussed in Appendix \ref{app:diffsketch}, choices of absorber and trap lengths, material thicknesses, and overlap fractions will have to be simulated and tuned to achieve meV-scale \Edep\ thresholds. It is also imperative that $n_0$ be sufficiently low in the absorber, otherwise the absorber may significantly elevate the trap background level.

\textbf{Vortices \& Quasiparticle Trapping: } Ref. \cite{wang2014vortex} discusses the impact of pinned (trapped) magnetic vortices, areas of quantized flux circulation. The local gap in these regions is driven to 0, and this sink-like behavior traps quasiparticles. Thus, pinned vortices in the QPD absorber could prevent quasiparticles from reaching the junctions. Ref. \cite{wang2014vortex} notices the geometrical dependence on pad shape for the vortex contribution, suggesting this might be one avenue to reduce their importance. Furthermore, very high magnetic permeability shielding could be used to prevent vortices from forming in the first place \cite{masuzawa2013magnetic}.

\textbf{Phonon Loss Simulations: } Any phonon losses directly degrade the energy resolution of the QPD. As discussed in Sec. \ref{sec:phonon}, literature values for \etaph\ vary widely between different experiments, making accurate determination of this parameter of vital importance. One promising avenue for accurate modeling is through use of bespoke phonon simulation software like G4CMP \cite{kelsey2023g4cmp}, with recent examples of phonon absorption simulations having been conducted for Kinetic Inductance Detectors \cite{martinez2019measurements} and qubits \cite{linehan2024estimating}. Experimental techniques to reduce phonon loss could include suspending the substrate via hanging wirebonds \cite{mcewen2022resolving} or in flip-chip designs where the ground plane and readout scheme are physically decoupled from the substrate \cite{rosenberg20173d,goupy2019contact}.

\section{Application to a rare-event search and conclusion}

To get a flavor of the science reach of QPDs, and with the performance curves of Sec. \ref{sec:resolutionandthreshold} in mind, we can consider their ability to probe for a light dark matter particle (mass $m_{\chi}$) elastically scattering off a much heavier target (i.e. a silicon atom with mass $m_{T}$) \cite{essig2022snowmass2021}. From basic kinematics, the maximum energy transfer in such a two-body collision, if $m_{T} \gg m_{\chi}$, is $E_{\rm max}\approx 2 m_{\chi}^2 v_{\chi}^2 / m_T$. The dark matter velocity $v_{\chi}$ is constrained to be the Milky-Way escape velocity of $\sim$600~km/s. Assuming all of $E_{\rm max}$ makes its way into the absorber (factoring in \etaph$\sim$30\%), a mixed Al/Hf device with a 75~meV deposit threshold, and no unexplained limiting background akin to a low-energy excess \cite{baxter2025low}, could probe down to $m_{\chi}\sim10$~MeV$c^{-2}$. Current experimental constraints for such elastically scattering light dark matter become weak at $m_{\chi}\sim30-100$~MeV$c^{-2}$ \cite{chang2025first, angloher2023results, collar2018search}. Thus, QPDs could provide close to an order-of-magnitude improvement in probing light dark matter parameter space.

In summary, we have outlined a scheme of using qubit-derived sensors for rare-event search particle physics applications, leveraging a rich literature of devices constructed for quantum information purposes. We have described the relevant physics and operating principles for these resonator coupled Quantum Parity Detectors along with preliminary calculations for potential noise sources and detector sensitivity. We have argued that such devices are intrinsically able to detect 10s of meV energy deposits and, given further R\&D into substrate-device couplings, should eventually be able sense single substrate-phonon events of a similar magnitude. QPDs could therefore be an excellent match to the challenge of detecting the lightest fermionic dark matter candidates. Ongoing fabrication efforts, based on designs presented in this paper, will hopefully experimentally verify the validity of the proposed schemes and calculations presented.

\begin{acknowledgments}
We thank SLAC and FNAL collaborators for insightful conversations regarding qubits, phonons, and quasiparticles. This research was primarily supported by National Science Foundation PHY Grant 2209581 and the Schwartz/Reisman Collaborative Science Program between Caltech and the Weizmann Institute of Science. This work was also supported in part by the Defense Advanced Research Projects Agency (DARPA) under the Quantum Sensing of Neutrinos program (QuSeN). KR was supported in part by a Caltech Prize Postdoctoral Fellowship and JP by the Caltech SURF Program.  This work was conducted at Washington University in St. Louis, California Institute of Technology, NASA's Jet Propulsion Laboratory, and the Weizmann Institute of Science.
\end{acknowledgments}

\newpage
\bibliography{bibs.bib}

\clearpage
\appendix

\section{Pulse Dynamics}\label{app:pulse}

Quasiparticle pulse dynamics in our thin film trap is described by the generation-recombination equation \cite{rothwarf1967measurement}, specialized to the case of injection of energy $P(t)$:
\begin{equation} \label{eq:nqpconservation}
    \frac{d\nqp}{dt} = \eta_{\rm ce}\eta_{\rm tr}\eta_{\rm pb,tr}\frac{P(t)}{V_{\rm tr}\Delta_{\rm tr}} - \frac{\nqp}{\tqp(\nqp)}, 
\end{equation}
where, in general, the lifetime in the second term is
\begin{equation}
\tqp(\nqp) = \frac{1}{2 R \nqp}
\end{equation}
where the recombination constant $R$ is material dependent. We assume that the injected power can be modeled as a simple exponential described by a single timescale:
\begin{align} \label{eq:phononpower}
\eta_{\rm ce} P(t) &= \frac{\Eabs}{\tau_{\rm inj}} e^{-t/\tau_{\rm inj}} \\
\tau_{\rm inj} &\approx \rm{max}(\tau_{\rm abs}, \tau_{qp,abs},\tau_{\rm tr}),
\end{align}
for, as mentioned in Sec. \ref{sec:devicephysics}, an injection timescale set by a combination of a phonon absorption timescale, a quasiparticle lifetime in the absorber, and a trapping timescale respectively.

The generation-recombination equation can be Taylor expanded for small a deviation $\dnqp$ from the background quasiparticle density $n_0$, allowing one to assume a fixed \tqp. That case, combined with our Eq. \ref{eq:phononpower} model of $P(t)$, yields the solution given by Eq.~\ref{eq:qp_pulse}. 

While this solution is not strictly valid for the case we consider here, $n_0 \ll \dnqp = \Nqpr/V$, we use this simple form with typical values of \tqp\ in order to capture the key elements of the dynamics. We believe these approximations are acceptable for this exploratory investigation while acknowledging that more precise modeling will be required for comparison to experimental demonstration data.

\section{Diffusion and Trapping from Absorber to Leads}\label{app:diffsketch}

\begin{figure}[!h]
    \includegraphics[width=0.48\textwidth]{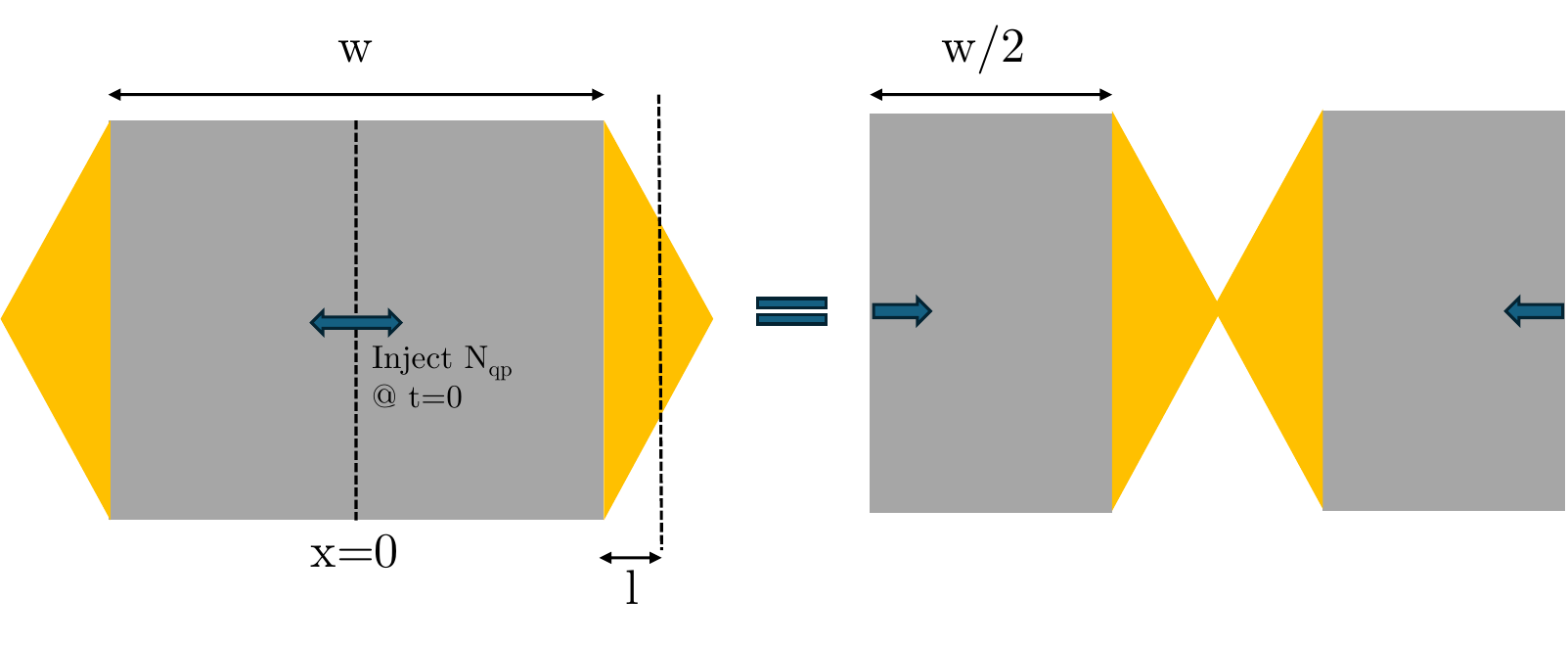}
    \caption{\label{fig:diffsketch} \textit{Left:} Sketch of a 1D diffusion model with $N$ quasiparticles injected in an absorber (gray) with length scale $w$ and trapped within a lower gap  region (gold) of length scale $l$. Non-trap boundaries are modeled as reflective. \textit{Right:} This is equivalent to a transmon geometry. Sketch not to scale.}
\end{figure}

We use the 1D analytic model of Angloher et al. \cite{angloher2016quasiparticle}, which describes the simplified geometry of Fig. \ref{fig:diffsketch}, to estimate and confirm the applicability of a trapping architecture for quasiparticles generated within a large absorber and funneled to a lower gap lead. 

The fraction of trapped quasiparticles, $\eta_{\rm tr}$, given the geometry and labeled parameters of the figure, is
\begin{equation} \label{eq:fracdiff}
    \eta_{\rm tr} = 2\frac{\rm{sinh}(\frac{\alpha}{2}) + \beta\rm{cosh}(\frac{\alpha}{2})}{(1+\beta^2)\rm{sinh}(\alpha) + 2\beta\rm{cosh}(\alpha)}
\end{equation}
for $\alpha=w/l_d$ where $l_d$ is a characteristic diffusion length dependent on absorber parameters (like film thickness). $\beta$ reflects a trapping probability, estimated in Ref. \cite{angloher2016quasiparticle} as $\beta=\tau_{\rm tr}/\tau_{\rm qp}$, for a quasiparticle trapping time and lifetime respectively. We modify this prescription, given our much smaller trap, as $\beta=\tau_{s}/\tau_{\rm dwell}$ for a quasiparticle scattering time (thereby losing energy below the absorber gap) and a characteristic dwell time within the trap. This parametrizes the absorption probability across the boundary, allowing for quasiparticles to effectively be reflected back into the absorber. 

\begin{figure}[!h]
    \includegraphics[width=0.40\textwidth]{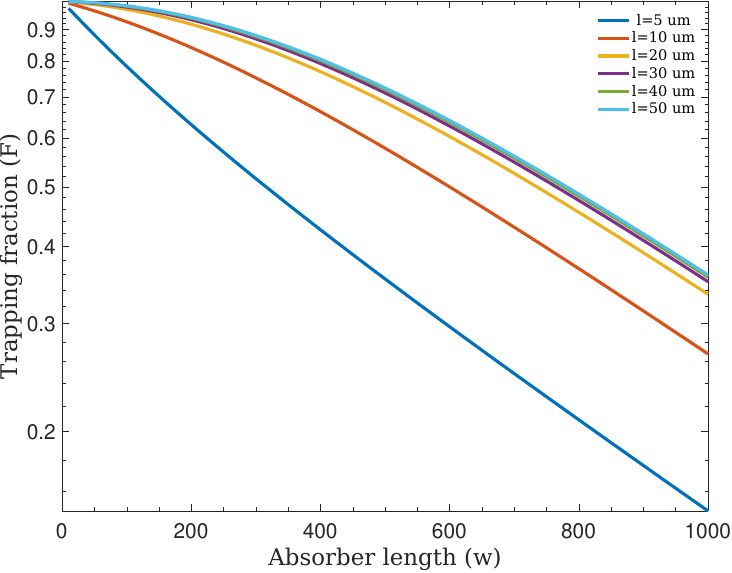}
    \caption{\label{fig:trapfraction} Fraction of trapped quasiparticles as computed from Eq. \ref{eq:fracdiff}.}
\end{figure}

We use $l_d = 300$~\um, an experimentally measured value for an Al absorber of thickness $500$~nm \cite{yen2015phonon}. $\tau_{s}\approx0.2\,\mu$s is computed using the formalism of Ref. \cite{kaplan1976quasiparticle}. The dwell time is estimated as $\tau_{\rm dwell}\sim l^2/D_{\rm trap}$, with trap length scale $l$ and trap diffusion constant $D_{\rm trap}\sim$2~cm$^2$s$^{-1}$, which has been scaled for a thinner 100~nm trap thickness \cite{yen2015phonon}. A resultant plot for the trapped quasiparticle fraction, for various trap lengths, is shown in Fig. \ref{fig:trapfraction}. We see that reasonable \bigo{0.1} values, in line with literature values \cite{moffatt2016two, yen2016quasiparticle} are achievable and can be plugged into our Eq. \ref{eq:edepres}.

The trapping efficiency depends on the quasiparticle transmission probability into the trap and a trapping timescale $\tau_{\rm tr}$ (the inverse of the rate at which a quasiparticle at $\Delta_{\rm abs}$ loses enough energy to be trapped, typically by phonon emission). Generally, only $\eta_{\rm tr}$ is experimentally measurable. For simplicity, we assume $\tau_{\rm tr} \ll \tau_{\rm inj}$ so it does not also need to be incorporated into Eq.~\ref{eq:qp_pulse}.

\section{Energy resolution of a sensor achieving maximal phonon coverage} \label{app:depositresolution}

As an anchor, we use the sensor absorbed energy resolution of $\sigma_{\rm ref}=6$~meV as per Fig. \ref{fig:E_abs_sweep} \textit{Right} for a Hafnium trap and junction of $V_{\rm ref}=1000$~\umthree. We envision a 1~cm$^2$ substrate, with 500~nm absorber films for phonon collection achieving 4\% surface coverage, as detailed in Sec. \ref{sec:phonon} resulting in some number $N_{\rm sens}$ of sensors required to fulfill the constraints of Eq. \ref{eq:abs}. 

We use the formalism of Appendix \ref{app:diffsketch} combined with our trapping relations of Sec. \ref{sec:phonon}, to ascertain that the resolution on energy deposited within the substrate will scale as

\begin{equation} \label{eq:edepres}
    \sigma_{\rm dep}= \frac{\sigma_{\rm ref}}{\eta_{\rm tr}\eta_{\rm pb,tr}\eta_{\rm ce}} \cdot \frac{\Delta_{\rm trap}}{\Delta_{\rm abs}} \cdot \bigg(\frac{V_{\rm trap}}{V_{\rm ref}}\bigg)^{1/2} \sqrt{N_{\rm sens}}
\end{equation}

\begin{figure}[!h]
    \includegraphics[width=0.40\textwidth]{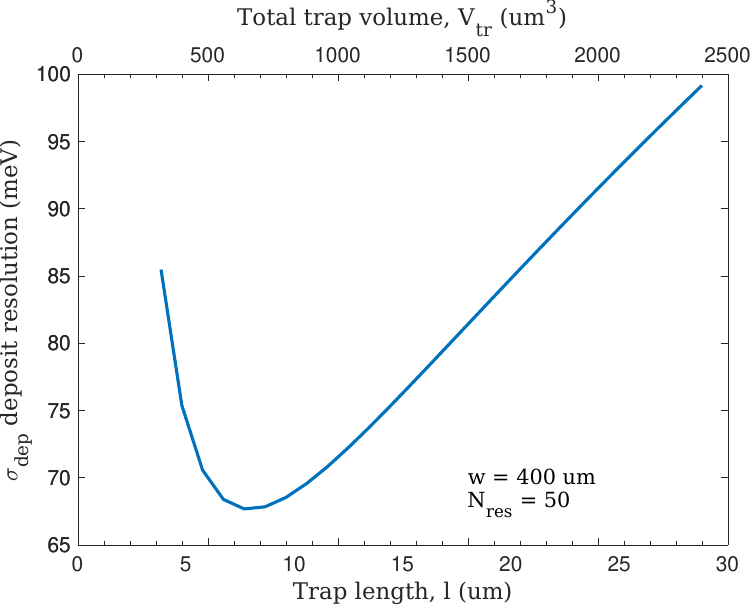}
    \caption{\label{fig:edepres} \textit{Top:} Expected resolution on energy deposited within the substrate, factoring in expected trapping and phonon efficiencies.}
\end{figure}

The results, plotted in Fig. \ref{fig:edepres}, support the conclusions of Sec. \ref{sec:estimates}. A more detailed 2D simulation, combining measured diffusion/absorption parameters in both trap and junction \textit{and} crucially factoring in the quiescent population $n_0$, will be required to determine an optimal geometry.

\section{Tunneling Rate}\label{app:closed} 

We first reproduce a derivation for the chemical potential shift (as found in Ref. \cite{shaw2008kinetics}) that sources the diffusion pressure for the quasiparticle tunneling. Following Ref. \cite{palmer2007steady} closely in this section, we start by noting that the total quasiparticle density in a volume will be a sum of the non-equilibrium and residual populations and can be given by
\begin{equation} \label{eq:totalpop}
    n_{\rm qp}+n_{\rm 0}=2\int_{\Delta}^{\infty}D_f\frac{\epsilon^{2}}{\sqrt{\epsilon^{2}-\Delta^{2}}}f_{ne}(\epsilon)d\epsilon,
\end{equation}
where $D_f=3\eta/2\epsilon_{F}$ is the normal metal density of states ($\eta$ is the conduction band electron density, and $\epsilon_{F}$ is the Fermi energy). The fractional term is the BCS density of states term for a superconductor. The factor of 2 accounts for hole-like and electron-like states. If there are only a small number of quasiparticles and the volume is in thermodynamic equilibrium, then the quasiparticle distribution is nominally given by the Fermi function
\begin{align} \label{eq:fermi_epsilon}
    f(\epsilon) & =\frac{1}{1+e^{\epsilon/k_{B}T}}  \\
    \epsilon &= \sqrt{\left(\frac{\hbar^{2}k^{2}}{2m}-\mu\right)^{2}+\Delta^{2}}, \nonumber
\end{align}
with $\epsilon$ denoting the energy of a quasiparticle excitation with wave vector $k$ in a superconducting volume. However, while the non-equilibrium quasiparticle population likely thermalizes rapidly ($\ll{\rm \mu s}$) to equilibrium temperature $T$, it is \textit{not} in chemical equilibrium. Thus, the chemical potential is shifted by $\delta \mu$ and the non-equilibrium quasiparticle distribution can instead be modeled by the Owen-Scalapino \cite{owen1972superconducting} form
\begin{equation} \label{eq:owen}
    f_{\rm ne}(\epsilon)=\frac{1}{1+e^{(\epsilon-\delta\mu)/k_{B}T}}.
\end{equation}

Ref. \cite{palmer2007steady} solves for the expression of the chemical potential shift in terms of the non-equilibrium quasiparticle density by re-expressing Eq. \ref{eq:totalpop} through
\begin{equation}
    n_{\rm qp}=2D_f\int_{\Delta}^{\infty}\frac{\epsilon^{2}}{\sqrt{\epsilon^{2}-\Delta^{2}}}(f_{\rm ne}(\epsilon)-f(\epsilon))d\epsilon,
\end{equation}
which, to lowest order in $k_BT/\Delta$, results in,
\begin{equation}
    \delta\mu\approx k_{B}T {\rm ln} \left[1+\frac{n_{\rm qp}}{\mathcal{N}}e^{\Delta/k_{B}T}\right],
\end{equation}
where $\mathcal{N}=D_f\sqrt{2\pi\Delta k_{B}T}$ is the density of available quasiparticle states. At low temperatures such that $n_{qp}e^{\Delta/k_{B}T}\gg\mathcal{N}$, per Ref. \cite{shaw2008kinetics}, the above expression reduces to
\begin{equation} \label{eq:chemshift}
    \delta\mu\approx\Delta+k_{B}T{\rm ln}\left[\frac{n_{\rm qp}}{\mathcal{N}}\right].
\end{equation}
In the case of asymmetric absorber and island, the terms in Eq. \ref{eq:chemshift} are indexed by their respective locations. Otherwise, they are equal for both pads.

We base our calculation of tunneling rates on Refs. \cite{catelani2011relaxation,catelani2014parity,martinis2009energy,serniak2019direct}, particularly the former two  . The quasiparticle tunneling Hamiltonian is
\begin{eqnarray} \label{eqn:tunnelhamiltonian}
    H_{T} = && \,\tilde{t}\sum_{l,r,s}\left[(u_{r}u_{l}-v_{r}v_{l}){\rm cos}\frac{\hat{\varphi}}{2}+i(u_{r}u_{l}+v_{r}v_{l}){\rm sin}\frac{\hat{\varphi}}{2}\right]\gamma_{r,s}^{\dagger}\gamma_{l,s} \nonumber \\
    &&+{\rm H.C.},
\end{eqnarray}
where $\tilde{t}$ is a tunneling matrix element, $\hat{\varphi}$ is the phase operator, $(u,v)$ are electron and hole occupation factors indexed by the left ($l$) and right ($r$) electrodes of the tunnel junction, and $\gamma$ are fermionic annihilation/creation operators ($s$ indexing spin). The tunneling (charge parity transition) rate from state $i$ to $j$ can then be computed from Eq. \ref{eqn:tunnelhamiltonian} using Fermi's Golden Rule. As worked out in Ref.\cite{catelani2011relaxation} this yields
\begin{equation} \label{eq:tunneling_rate_full}
    \Gamma_{ij}=\frac{16E_{J}}{\pi\hbar}\left(|\langle j|{\rm cos}\frac{\hat{\varphi}}{2}|i\rangle|^{2}S_{ij}^{-}+|\langle j|{\rm sin}\frac{\hat{\varphi}}{2}|i\rangle|^{2}S_{ij}^{+}\right),
\end{equation}
where, in the low-energy regime, the tunneling rate has been factorized into two components: the matrix elements account for qubit state transitions while the $S_{ij}^{\pm}$ are ``spectral functions" that handle initial state occupancy and final state blocking. In this detector concept, we do not rely on state changes and hence primarily work only with the ground state ($i=0$, $j=0$, $\delta\varepsilon_{ij}\sim0$). Further, the electrodes are equivalent ($\epsilon_{l,r}\to\epsilon$) with equal chemical potential shifts. The only relevant spectral function is given by
\begin{equation} \label{eq:spectral}
    S_{00}^{\pm}=\frac{1}{\Delta}\int_{\Delta}^{\infty}d\epsilon f_{ne}(\epsilon)[1-f_{ne}(\epsilon)]\frac{\epsilon^2}{\epsilon^{2}-\Delta^{2}}\left(1\pm\frac{\Delta^{2}}{\epsilon^{2}}\right).
\end{equation}

For the transmon case, where $\chi = E_C/E_J << 1$, the cos and sin terms are shown in \cite{catelani2014parity} to reduce to
\begin{align} \label{eq:anglefactors}
    |\langle0,p|{\rm cos}\frac{\hat{\varphi}}{2}|0,p\rangle| \approx &1-\frac{1}{2}\sqrt{\frac{E_{C}}{8E_{J}}}-\frac{3}{64}\frac{E_{C}}{E_{J}}\sim0.9\approx1 \nonumber \\
    |\langle0,p|{\rm sin}\frac{\hat{\varphi}}{2}|0,p\rangle| \approx & |{\rm sin}(2\pi n_{g})|\left(\frac{2}{3}\right)^{2/3} \nonumber \\
    &\times \Gamma \left(\frac{1}{3}\right)\left(\frac{E_{C}}{8E_{J}}\right)^{\frac{1}{6}}\frac{\epsilon_{i}}{\omega_{p}}\approx 0,
\end{align}
where $\Gamma$ in the last expression is the gamma function.

Inserting Eqs. \ref{eq:spectral} and \ref{eq:anglefactors} into \ref{eq:tunneling_rate_full} with the help of \ref{eq:chemshift}, we obtain:
\begin{align}
    \Gamma_{00}&=\frac{32E_{J}}{h}\frac{1}{\Delta}\int_{\Delta}^{\infty}f_{ne}(1-f_{ne})d\epsilon \nonumber \\
    &=\frac{16E_{J}}{h\Delta}\int_{\Delta}^{\infty}\frac{1}{1+{\rm cosh}(\frac{\epsilon-\delta\mu}{k_{B}T})}d\epsilon \nonumber \\
	&=\frac{16E_{J}k_{B}T}{h\Delta}\frac{1}{1+e^{(\Delta-\delta\mu)/k_{B}T}} \nonumber \\
	&\approx\frac{16E_{J}k_{B}T}{h\Delta}\frac{1}{1+\frac{\mathcal{N}}{n_{\rm qp}}} \approx\boxed{\frac{16E_{J}k_{B}T}{\mathcal{N}\Delta h}n_{\rm qp}\equiv Kn_{\rm qp}},
\end{align}
where we have assumed \nqp $\ll\mathcal{N}$ in arriving at the final expression. Eq. \ref{eq:K_equation} is thus proven.

\section{Diffusion \& Tunneling Simulations within the Leads}\label{app:diffusion}

We have made the assumption that the \nqp\ injected into the neighborhood local to the junction thermalizes rapidly and is uniform throughout the trap, motivating the use of a chemical shift. In reality, quasiparticle deposits in the trap take time to diffuse to the junction and tunnel across. We briefly investigated the validity of this prescription, following Ref. \cite{wang2014vortex}, by modeling quasiparticle diffusion and tunneling as,
\begin{eqnarray}
    \label{eq:qp_diff_eq}
    \frac{\partial u(x,y,t)}{\partial t} = && D\nabla^2 u(x,y,t) - \frac{K}{V_{\rm trap}} u(x,y,t) \delta (\vec{r} - \vec{r}_{j}) \nonumber \\
    && - \frac{1}{\tau_{\rm qp}} u(x,y,t).
\end{eqnarray}
We assume reflective Neumann boundary conditions and have reduced the problem to two-dimensions (as film thickness $\ll$ phonon absorber size). Tunneling is modeled as a delta-function sink term. As in Appendix \ref{app:pulse}, we assume a fixed quasiparticle lifetime, in this case taking $\tau_{\rm qp}=1$~ms as a typical value at low quasiparticle density \cite{zmuidzinas2012superconducting}. We use a value of $K \approx 3\ \si{kHz\ \mu m^3}$ similar to Sec. \ref{sec:estimates}. $V_{\rm tr}$ is the volume of the junction, and trap dimensions are taken to match this volume, with investigated geometries ranging from square to aspect ratios of $\sim$10. $r_j$ is the location of the junction. $u(x,y,t)$ is normalized to unity such that it can be interpreted as the PDF of a quasiparticle in the absorber. We set $D = 2~\si{cm}^2~\si{s}^{-1}$, assuming that the trap region diffusion is similar to Aluminum. Literature values range from $D=22.5 -34~\si{cm}^2~\si{s}^{-1}$ \cite{hsieh1968diffusion}. The latter comes from a measurement on 20~nm thick aluminum (thinner than our proposed traps), performed by a subset of authors of this paper, where they extracted the diffusion constant through its relationship to the critical superconducting field $H_c$ as per Ref. \cite{gordon1986electron}). The choice of a much smaller diffusion constant in our simulations comes from certain reports indicating poor diffusion for thin ($<100$~nm) films.\cite{fink2022gram,yen2015phonon, yen2016quasiparticle}.

The instant diffusion limit ($D \rightarrow \infty$) solution of Eq. \ref{eq:qp_diff_eq} is equivalent to having the junction be the entire volume of the trap: $\dot{u}(t) = - Ku(t)/V_{\rm tr}  - u(t)/\tau_{\rm qp}$. The normalized absorber quasiparticle density thus becomes $u(t) = e^{-t/\tau_t} e^{-t/\tau_{\rm qp}}$, where $\tau_t = V/K$ is the tunneling time. We compare this regime to one with the measured diffusion constant and simulate solutions to Eqn~(\ref{eq:qp_diff_eq}) using standard finite-difference techniques. From this process, we ascertain a few things: 1) agreement between the simulated solution and the instant diffusion limit justify our use of a uniform quasiparticle density, $n_{\text{qp}}$, and a chemical potential, $\delta \mu$, across the trap; 2) quasiparticles become uniformly distributed over the trap regardless of the initial location of the energy deposition (including uniform injection from the sidewall opposite the junction) and aspect ratio; 3) $\tau_{\text{qp}} \ll \tau_t$ so recombination dominates over tunneling. In particular, we can view the tunneling rate as a measure of the trap quasiparticle density; 4) Eq. \ref{eq:qp_pulse}, in spite of diffusion, is a good description of the uniform, recombination-dominated quasiparticle density. In totality, the results obtained from the simulation are consistent with the conclusions presented in Sec. \ref{sec:devicephysics}.

\section{Absorber and Trap Balance}\label{app:detailed}
Here we show that a source of non-thermal background quasiparticles in the absorber can lead to greatly elevated quasiparticle densities in the trap. A simplified model of the steady-state quasiparticle population within the trap and absorber can be written as set of coupled differential equations. The density in the absorber can be expressed as

\begin{equation}
    \frac{dn_{\rm qp, abs}}{dt} = \frac{1}{V_{\rm abs}}(\Gamma_{\rm gen}^{\rm abs} - \Gamma_{\rm rec}^{\rm abs}) - p\frac{V_{\rm tr}}{V_{\rm abs}}\frac{n_{\rm qp, abs}}{\tau_{\rm tr}},
\end{equation}

where $\Gamma_{\rm gen,rec}^{\rm abs}$ are quasiparticle generation and recombination (GR) terms, inclusive of any thermal or pair-breaking radiation contribution. The last term represents the fractional loss into the trap with an assumed trapping timescale $\tau_{\rm tr}$ related to the phonon emission timescale (among other considerations). We have introduced a transmission probability in the last term, $p$, such that with perfect transmission quasiparticles dwell within the trap for time in proportion to the volume of the trap vs. the absorber. 

\begin{figure}[!h]
    \includegraphics[width=0.40\textwidth]{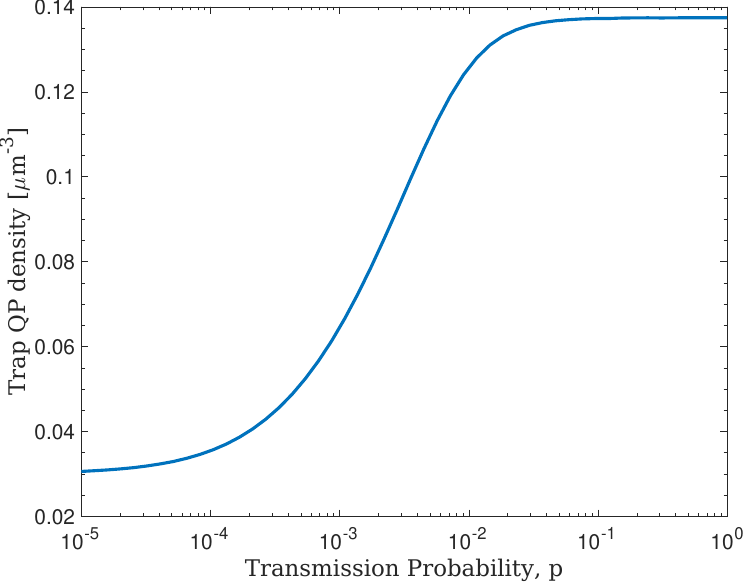}
    \caption{\label{fig:transprob} Trap quasiparticle density as a function of transmission probability from the absorber, for values given in text. As the capture probability increases, the more elevated the trap density can become at steady-state, directly affecting detector sensitivity.}
\end{figure}

The corresponding equation for the trap is

\begin{eqnarray}
    \frac{dn_{\rm qp, tr}}{dt} = && \frac{1}{V_{\rm tr}}(\Gamma_{\rm gen}^{\rm tr} - \Gamma_{\rm rec}^{\rm tr}) + p \frac{n_{\rm qp, abs}}{\tau_{\rm tr}}\eta_{\rm pb}\frac{\Delta_{\rm abs}}{\Delta_{\rm tr}},
\end{eqnarray}
where the first term corresponds again to quasiparticle generation and recombination and the second term is the source term due to trapping, copied over from the first equation and accounting for the different volume. While experimental input will be required to correctly constrain such a model, we can use reasonable assumptions to evaluate the dynamics. We assume elevated but still very low quasiparticle densities of approx. $\mathcal{O}$(0.01)~um$^{-3}$ in the trap, a $\mathcal{O}$(1)~ms quasiparticle lifetime in the absorber, a $\mathcal{O}$(100)~$\mu$s lifetime within the trap, a $\mathcal{O}$(100)~ns trapping timescale, and a constant injection of quasiparticles into the absorber that would, at steady-state, result in an absorber density of $\sim$0.05 qp$\cdot$\umminusthree. We can solve the system of equations, with the resulting trap density vs. transmission probability shown in Fig. \ref{fig:transprob}. We establish our conclusion from Sec. \ref{sec:estimates}: at steady state, and as the transmission probability improves, the trap quasiparticle density can be greatly elevated from its quiescent value and can be many times larger.

\end{document}